\documentclass[amsfonts,showpacs,tightenlines,aps,12pt,floatfix,nofootinbib]{revtex4}
\usepackage{bm}
\usepackage{epsfig}
\usepackage{float}
\begin{document}

\thispagestyle{empty}

\title{NNLO low-energy constants from flavor-breaking 
chiral sum rules based on hadronic $\tau$-decay data}

\author{Maarten Golterman}
\email[]{maarten@stars.sfsu.edu}
\affiliation{Department of Physics and Astronomy,
San Francisco State University, San Francisco, CA 94132, USA}
\author{Kim Maltman}
\email[]{kmaltman@yorku.ca}
\affiliation{Department of Mathematics and Statistics, York University,
4700 Keele St., Toronto, ON CANADA M3J 1P3}
\altaffiliation{CSSM, Univ. of Adelaide, Adelaide, SA 5005 AUSTRALIA}
\author{Santiago Peris}
\email[]{peris@ifae.es}
\affiliation{Department of Physics, Universitat Aut\`onoma de Barcelona
\\ E-08193 Bellaterra, Barcelona, Spain}

\begin{abstract}
Using spectral data from non-strange and strange hadronic $\tau$ decays,
flavor-breaking chiral sum rules involving the flavor $ud$
and $us$ current-current two-point functions are constructed and
used to determine the $SU(3)$ NNLO low-energy constant
combinations $C_{61}^r$, $C_{12}^r+C_{61}^r+C_{80}^r$
and $C_{12}^r-C_{61}^r+C_{80}^r$. 
The first of these determinations updates the results of an earlier 
analysis by D\"urr and Kambor, while the latter two are new.
The error on the $C_{12}^r+C_{61}^r+C_{80}^r$ is particularly small. 
Comparisons are made to model estimates for these quantities. The role 
of the third combination in significantly improving the determination 
of the NLO low-energy constant $L_{10}^r$ from NNLO analyses of the 
flavor $ud$ V-A correlator is also highlighted.
\end{abstract}

\pacs{12.39.Fe,11.55.Hx,13.35.Dx}

\maketitle

\section{\label{intro}Introduction}
Chiral perturbation theory (ChPT) provides a means of implementing, in the 
most general way, the constraints on low-energy processes of the symmetries 
of QCD~\cite{weinberg79,gl84,gl85}. The 
effects of resonances, and other heavy degrees of freedom, are encoded in the 
low-energy constants (LECs) which appear in the resulting effective chiral
Lagrangian multiplying those operators allowed by these constraints. In
the even-intrinsic-parity sector, at next-to-leading order (NLO) in the 
chiral counting, the $SU(3)\times SU(3)$ Lagrangian 
involves $10$ in-principle-measurable LECs, the $L_k$ introduced in 
Ref.~\cite{gl85}. The next-to-next-to-leading order (NNLO) form was 
first considered in Ref.~\cite{fs96}, and a reduced minimal set of operators 
subsequently found in Refs.~\cite{bce99}. The minimal NNLO $SU(3)$ form 
involves $94$ additional LECs, $4$ in contact and $90$ in non-contact terms. 
In what follows, we work with the dimensionful versions, $C_k$, of the NNLO 
LECs introduced in Refs.~\cite{bce99}. 

To make the NNLO chiral Lagrangian fully predictive, existing determinations 
of the $L_k$ must be supplemented with model-independent experimental 
and/or theoretical determinations of the $C_k$. To date a limited number of 
such determinations exist. 

First attempts at obtaining what is now called 
$C_{61}$ were made in Refs.~\cite{kms94,gk95,mw9599}, with a more robust 
chiral sum rule determination, involving the flavor-breaking (FB) $ud$-$us$ 
vector current correlator, obtained in Ref.~\cite{dk00}. $C_{12}$ and 
the combination $C_{12}\, +\, C_{34}$ were determined via 
phenomenological~\cite{jop04,bp07} and lattice~\cite{bp10} analyses 
of the scalar $K\pi$ form factor, and $C_{14}\, +\, C_{15}$ and
$C_{15}\, +\, 2C_{17}$ from analyses of the quark-mass-dependence of 
lattice data for $f_K/f_\pi$~\cite{bp10,emn10,emn13} (some aspects 
of these latter analyses employing, in addition, large-$N_c$ arguments). 
Generally less precise 
constraints on the combinations $C_{88}-C_{90}$, $2C_{63}\, -\, C_{65}$ 
and $6C_{12}\, +\, 2C_{63}\, +\, 2C_{65}\, +\, 3C_{90}$ were obtained from 
analyses of the charged $\pi$ and $K$ electromagnetic form 
factors~\cite{bt02}, and on the combinations $C_{12}\, +\, 2C_{13}$, $C_{13}$ 
and $C_{12}\, +\, 4C_{13}$ from analyses of the curvature of the 
$\pi$ and strangeness-changing $K\pi$ scalar form factors~\cite{bd03}.
An overconstrained (but, with current data, not yet fully self-consistent) 
determination of the set $C_{1-4}$ was also made~\cite{km06}, using a 
combination of four of the subthreshold 
coefficients of the $\pi K$ scattering amplitudes determined in 
Ref.~\cite{bdm03} and two of the low-energy $\pi\pi$ scattering 
parameters determined in Ref.~\cite{cgl01}. The four remaining 
$\pi\pi$ scattering parameters and six remaining $\pi K$ subthreshold 
coefficients provide ten additional constraints 
on the $24$ NNLO LECs $C_{5-8},\, C_{10-17},\, C_{19-23},\, C_{25},\,
C_{26}$ and $C_{28-32}$~\cite{km06}. 
Finally, $C_{87}$ has been determined from analyses of the 
light-quark V-A current-current correlator~\cite{gapp08,bgjmp13}.

In the absence of clean theoretical and/or data-based 
determinations, it is common to use estimates of the $C_k$ obtained
in model-dependent approaches. One such strategy is to
extend the resonance ChPT (RChPT) approach~\cite{rchpt} (often
held to work well in estimating NLO LECs~\cite{pichrchptreview}) to 
NNLO~\cite{resonance_C_k}. This approach typically employs, in addition
to long-distance chiral constraints, short-distance QCD and large-$N_c$ 
constraints. Evidence exists that at least some $1/N_c$-suppressed 
LECs cannot be neglected~\cite{ghi07,bj11} (we will comment below on 
another such piece of evidence). A second approach to estimating the 
$C_k$, in a large-$N_c$ gauge-invariant non-local quark model framework, 
was presented in Ref.~\cite{jzlw09}. Comparisons performed in 
Refs.~\cite{km06,jzlw09} between predicted $C_k$ values and those known 
from experiment expose some shortcomings in both approaches.

In light of this situation, additional model-independent NNLO LEC 
determinations are of interest, first as part of the ongoing long-term 
program of pinning down the parameters of the low-energy effective 
Lagrangian, and second, as a means of further testing, and constraining, 
models used to estimate additional as-yet-undetermined LECs. In this paper, 
we update the earlier determination of $C_{61}$~\cite{dk00} and provide
a new high-precision determination of the combination $C_{12}+C_{61}+C_{80}$.
With input for $C_{12}$ from other sources (such as those noted above) this 
yields also a determination of $C_{80}$. A direct determination of the 
combination $C_{12}-C_{61}+C_{80}$ which, with the $1/N_c$-suppressed 
combination $C_{13}-C_{62}+C_{81}$, is needed to complete the determination 
of the NLO LEC $L_{10}$ from an NNLO analysis of the low-energy behavior of 
the light quark V-A correlator~\cite{gapp08,bgjmp13} is also
obtained. Combining this determination with the continuum light-quark 
V-A correlator analysis of Ref.~\cite{bgjmp13} and lattice analysis of 
Ref.~\cite{boylelatt13vma} turns out to make possible a high-precision 
($\sim 10\%$) determination of $L_{10}$. This level of precision
requires careful consideration of the LEC combination 
$C_{13}-C_{62}+C_{81}$ which, though nominally subleading in $1/N_c$, 
turns out to have a non-zero value comparable to that of
the non-$1/N_c$-suppressed combination $C_{12} - C_{61} + C_{80}$
(although with large errors)~\cite{bgjmp13}. This non-zero value has 
a non-trivial impact on the determination of $L_{10}$, shifting the
magnitude of the result by $15\%$ compared to what is obtained if
$C_{13}-C_{62}+C_{81}$ is instead set to zero
on the grounds of its $1/N_c$ suppression~\cite{boylelatt13vma}.

The rest of the paper is organized as follows. In Section~\ref{imsr}, we 
introduce, and give the explicit forms of, the chiral sum rules 
to be employed. In Section~\ref{input} the experimental, NLO LEC, and OPE
inputs to these sum rules are specified. Section~\ref{results} contains 
the results and a comparison to model predictions for the LEC combinations 
in question. Section~\ref{summary}, finally, contains a brief summary.
Details of the OPE contributions and errors are gathered in an
appendix.

\section{\label{imsr}The flavor-breaking chiral sum rules}
The key objects for the analysis described in this paper 
are the flavor $ij=ud,us$ vector ($V$) and axial vector ($A$) 
current-current two-point functions, $\Pi_{V/A}^{\mu\nu}$,
and their spin $J=0,1$ components, $\Pi_{ij;V/A}^{(J)}$.
These are defined by
\begin{eqnarray}
\Pi_{ij;V/A}^{\mu\nu}(q^2)&&\, \equiv\, i\int\, d^4x\, e^{iq\cdot x}
\langle 0\vert T\left( J_{ij;V/A}^\mu (x) J_{ij;V/A}^{\dagger\, \nu} (0)\right)
\vert 0\rangle
\nonumber\\
&&\,=\, \left( q^\mu q^\nu - q^2 g^{\mu\nu}\right)\, \Pi^{(1)}_{ij;V/A}(Q^2)
\, +\, q^\mu q^\nu\, \Pi_{ij;V/A}^{(0)}(Q^2)\ ,
\label{mink2pt}\end{eqnarray}
where $J_{ij;V/A}$ are the standard flavor $ij$ $V/A$ currents, and 
$Q^2\, =\, -q^2\, =\, -s$. $\Pi^{(0,1)}_{ij;A}$ individually have
kinematic singularities at $Q^2=0$, but their sum, $\Pi^{(0+1)}_{ij;A}$, 
and $s\Pi^{(0)}_{ij;A}$ are both kinematic-singularity-free.
The associated spectral functions, $\rho^{(J)}_{ij;V/A}(s)\, =\,
{\rm Im}\, \Pi^{(J)}_{ij;V/A}(s)/\pi$, are accessible experimentally
through the normalized differential distributions, $dR_{ij;V/A}/ds$, 
\begin{eqnarray}
R_{ij;V/A}\, &&\equiv\, \Gamma [\tau^- \rightarrow \nu_\tau
\, {\rm hadrons}_{ij;V/A}\, (\gamma )]/ \Gamma [\tau^- \rightarrow
\nu_\tau e^- {\bar \nu}_e (\gamma)]\, ,
\end{eqnarray}
measured in flavor $ij$ $V$- or $A$-current-induced hadronic $\tau$ decays.
Explicitly~\cite{tsai}
\begin{eqnarray}
&&{\frac{dR_{ij;V/A}}{ds}}\, =\, {\frac{12\pi^2\vert V_{ij}\vert^2 S_{EW}}
{m_\tau^2}}\, \left[ w_\tau \left( y_\tau \right)\, \rho_{ij;V/A}^{(0+1)}(s)\,
-\, w_L \left( y_\tau \right)\, \rho_{ij;V/A}^{(0)}(s) \right]\ ,
\label{basictaudecay}\end{eqnarray}
with $y_\tau = s/m_\tau^2$, $w_\tau (y)=(1-y)^2(1+2y)$, $w_{L}(y)=2y(1-y)^2$, 
$S_{EW}$ a known short-distance electroweak correction~\cite{erler},
and $V_{ij}$ the flavor $ij$ CKM matrix element. The dominant contributions
to $\rho_{ud,us;A}^{(0)}(s)$ are the accurately known, chirally
unsuppressed, $\pi$ and $K$ pole terms. The remaining $J=0$ $V/A$ spectral
contributions are proportional to $(m_i\mp m_j)^2$, hence numerically 
negligible for $ij=ud$. $\rho^{(0+1)}_{ud;V+A}(s)$ is thus determinable 
directly from the non-strange differential decay distribution. For $ij=us$, 
phenomenological determinations strongly constrained by the known strange 
quark mass are available for the small continuum scalar 
$\rho_{us;V}^{(0)}(s)$~\cite{jopss} and pseudoscalar 
$\rho_{us;A}^{(0)}(s)$~\cite{kmsps} contributions in the region 
$s<m_\tau^2$ relevant to hadronic $\tau$ decays. With the contributions
proportional to $w_L \left( y_\tau \right)\, \rho_{us;V/A}^{(0)}(s)$ in 
Eq.~(\ref{basictaudecay}) thus fixed, $\rho^{(0+1)}_{us;V+A}(s)$ can be
determined from the strange differential decay distribution. The $V/A$ 
separation for the $ud$ and $us$ cases will be discussed further in the 
next section.
\begin{figure}[H]
\caption{\label{f1}The contour underlying the chiral sum rules of
Eq.~(\ref{imsrbasic})}
\centering
\includegraphics[width=2in]{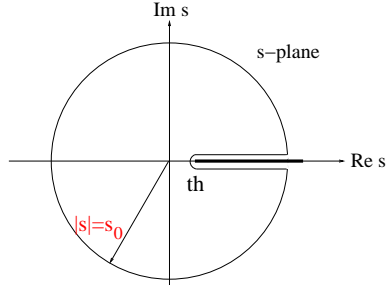}
\end{figure}

Given a correlator, $\Pi (Q^2=\, -s)$, free of kinematic 
singularities, and the corresponding spectral function, $\rho (s)$,
application of Cauchy's theorem to the contour shown in Fig.~\ref{f1} yields 
the inverse moment (chiral) finite energy sum rule (IMFESR) 
relation, valid for any choice of weight function, $w(s)$, analytic
in the region of the contour,
\begin{eqnarray}
&&w(0)\, \Pi (0) \, =\, {\frac{1}{2\pi i}}\,\oint_{\vert s\vert = s_0} ds\, 
{\frac{w(s)}{s}}\, \Pi (Q^2)\ +\ \int_{th}^{s_0}ds\, 
{\frac{w(s)}{s}}\,\rho (s)\ ,
\label{imsrbasic}\end{eqnarray}
where $th$ is the relevant physical threshold.
We will work below with values of $s_0$ and FB correlator combinations, 
$\Pi (Q^2)$, such that the $\rho (s)$ needed on the RHS are accessible 
from hadronic $\tau$-decay data. A determination of the combination of 
LECs occurring in the chiral representation of $\Pi (0)$ is then obtained 
by inputting the chiral representation on the LHS and evaluating both terms 
on the RHS. For large enough $s_0$, the first term on the RHS can be 
evaluated using the OPE representation of $\Pi (Q^2)$, while, for 
$s_0<m_\tau^2$, the second term can be evaluated using experimental 
spectral data. Previous sum rule studies have, however, found that,
even for $s_0\sim 2-3\ {\rm GeV}^2$, integrated duality violations 
(OPE breakdown) can be sizeable for $w(s)$ which are not
zero at the timelike point $s=s_0$ on the contour~\cite{kmdv,dv71,dv72}.
We thus further restrict our attention to $w(s)$ satisfying $w(s_0)=0$. 

We will consider two choices for the weight $w(y)$, $y=s/s_0$:
\begin{eqnarray}
w_{DK}(y)&=&(1-y)^3\left(1+y+\frac{1}{2}y^2\right)=
1-2y+\frac{1}{2}\left[y^2+y^3+y^4-y^5\right] \ ,\nonumber\\
\hat{w}(y)&=&(1-y)^3 \ .
\label{weightdefns}\end{eqnarray}
The first of these was considered in Ref.~[9].
Both weights satisfy $w(0)=1$ and are ``triply pinched'' ({\it i.e.}, have 
a triple zero at $s=s_0$), strongly suppressing duality violating 
contributions to the first term on the RHS of Eq.~(\ref{imsrbasic}). 
An additional advantage of the triple zero is the suppression of 
contributions to the weighted spectral integrals (the second term on 
the RHS of Eq.~(\ref{imsrbasic})) from the high-$s$ part of the spectral 
functions, where the $us$ data currently available suffers from low 
statistics and large $V/A$ separation uncertainties~\cite{dk00}. The 
strong suppression at large $s$ for these weights is thus doubly beneficial 
to the goal of this article, which is to determine as accurately as 
possible the LHSs, $w(0)\Pi(0)=\Pi(0)$, of Eq.~(\ref{imsrbasic}) for 
various FB combinations, $\Pi$, of the $ud$ and $us$ $V$ and $A$ 
correlators (see Eq.~(\ref{DeltaPidefns}) below).{\footnote{The 
motivation for the choice of weights here is to be contrasted with 
that in Refs.~\cite{dv71,dv72}, in which the principal aim was a 
precision determination of $\alpha_s(m_\tau^2)$ from the non-FB
$ud$ $V$ and $A$ correlators. The need for high ($\sim 1\%$ or less)
precision on the theoretical side of the sum rules employed in that case 
favors a restriction to weights of lower degree, which minimize the 
number of $D\ge 6$ OPE condensates that need to be fit to data, but have
less pinching than those in the present case. As the level of
pinching decreases, the possibility for significant integrated
duality violations increases.
As explained in detail in Ref.~\cite{dv71}, 
the inclusion of unpinched weights in that analysis allowed for
the modelling and constraining of these contributions, providing a 
means for investigating quantitatively the level of integrated 
duality violation not only in that case but also in earlier 
pinched-weight analyses. Here, the situation is different, as the
use of triply pinched polynomials $w(y)$ in the full weights,
$w(y)/s$, used to access the low-$s$ physics of interest
to us, not only strongly suppresses duality violating contributions,
as already noted above, but also significantly reduces the errors on 
the weighted $us$ spectral integrals. The additional 
$1/s$ factor in the weights further helps by reducing the maximum
dimension of OPE condensates which have to be considered in the
analysis. The final important difference between the present
case and that of the $\alpha_s$ analysis is the precision
required for the OPE contributions. Here, OPE contributions
turn out to play a smaller numerical role, greatly reducing the 
level of precision required for the evaluation of these contributions.}}

The value of $\Pi(0)$ in Eq.~(\ref{imsrbasic}) should, of course,
be independent both of $s_0$ and the choice of weight, $w(s)$.
Verifying that these independences are in fact realized provides
non-trivial tests of the self-consistency of the theoretical and 
spectral input to the analysis.

In the rest of the paper, we concentrate on IMFESRs involving one of 
the three choices, $T=V$, $V\pm A$, of the FB $ud-us$ combinations of $J=0+1$
$V$ and $A$ correlators,
\begin{equation}
\Delta\Pi_T\equiv\Pi^{(0+1)}_{ud;T}-\Pi^{(0+1)}_{us;T}\ .
\label{DeltaPidefns}\end{equation}
The corresponding spectral functions are denoted $\Delta\rho_T$.
Versions of the $T=V\pm A$ correlator and spectral-function combinations
having their $\pi$ and $K$ pole contributions subtracted will be denoted
by $\Delta\overline{\Pi}_{V\pm A}$ and $\Delta\overline{\rho}_{V\pm A}$. 
The restriction to the $J=0+1$ combination is predicated on the very bad 
behavior of the OPE representation of $\Pi^{(0)}_{ij;V/A}(Q^2)$ on the 
contour $|Q^2|=s_0$ for all $s_0$ accessible using $\tau$ decay 
data~\cite{kmtauprob,kmlong}.

It is worth commenting on the differences in the OPE contributions
for the two weights $w_{DK}$ and $\hat{w}$. We focus here on
$D\ge 4$ contributions ($D=2$ contributions will be discussed in more 
detail later, as will $D=4$ contributions higher order in $\alpha_s$). 
For a general polynomial $w(y)=\sum_{m=0} a_my^m$, writing 
$[\Delta\Pi_T(Q^2)]^{OPE}_{D\ge 4}$ in the form 
$\sum_{k\ge 2}C^T_{2k}/Q^{2k}$, with $C^T_D$ an effective 
dimension-$D$ condensate, the integrated $D\ge 4$ OPE contributions 
to the RHS of Eq.~(\ref{imsrbasic}) become
\begin{equation}
\frac{1}{2\pi  i}\oint_{|s|=s_0}ds\,\frac{w(s/s_0)}{s}\,
[\Delta\Pi_T(Q^2)]^{OPE}_{D\ge 4}=\sum_{k\ge 2}(-1)^ka_k\,
\frac{C^T_{2k}}{s_0^k}\ ,
\label{higerdopeint}\end{equation}
up to logarithmic corrections suppressed by additional powers of $\alpha_s$. 
Since $\hat{w}$ has degree three, only OPE contributions up to $D=6$ 
contribute, if we ignore the $\alpha_s$-suppressed logarithmic 
corrections. In contrast, $w_{DK}$, which has degree five, produces 
contributions at leading order in $\alpha_s$ up to $D=10$. In this sense, 
$\hat{w}$ is preferred over $w_{DK}$, since the latter involves
additional unknown or poorly known $D=8$ and $D=10$ condensates.
For $w_{DK}$, these contributions are expected to be small, partly because 
the coefficients $a_m$, $m=4,\ 5$ are small enough to avoid unwanted
enhancements, and partly because of the
$1/s_0^{D/2}$ suppression of such higher $D$ contributions for
$s_0\gg\Lambda_{QCD}^2$, but this expectation can (and should) be tested. 
Performing the IMFESR analysis for a range of $s_0$ provides a 
means of doing so and, in fact, provides a test of the 
reliability of any approximations employed in evaluating the RHS,
including also the neglect of integrated duality violations. The
$s_0$ dependence of the RHS of Eq.~(\ref{imsrbasic}) will be considered 
for the combinations of Eq.~(\ref{DeltaPidefns}) for both weights 
in Sec.~\ref{results} below.

An advantage of the weight $w_{DK}$ over $\hat{w}$ is that the $D=2,\ 4$ 
contributions to the integral in Eq.~(\ref{higerdopeint}) are better 
behaved for $w_{DK}$. In addition, as can be seen from 
Eq.~(\ref{higerdopeint}), since $a_2^{\hat{w}}/a_2^{w_{DK}} = 6$,
the leading-order $D=4$ contribution, and associated error, are
both a factor of $\sim 6$ larger for $\hat{w}$. Analogous, though
somewhat smaller, enhancement factors, 
$\vert a_1^{\hat{w}}/a_1^{w_{DK}}\vert \, =\, 3/2$ and
$\vert a_3^{\hat{w}}/a_3^{w_{DK}}\vert \, =\, 2$ 
are operative for the leading order $\hat{w}$ $D=2$ and $D=6$ 
contributions and errors. $D=2,\ 4$ and $6$ contributions, and hence total 
OPE errors, are thus significantly larger for $\hat{w}$ than for $w_{DK}$
from this effect alone. The nominal convergence of the known terms in 
the integrated $D=2$ OPE series is also significantly slower for $\hat{w}$. 
The larger OPE errors turn out to produce total errors for $\hat{w}$ 
which are similar to those for $w_{DK}$ ({\it cf}. Table~\ref{table2} 
in Sec.~\ref{results}). The differences in the relative sizes of spectral 
integral and OPE contributions to the RHSs of the $w_{DK}$ and $\hat{w}$ 
IMFESRs also means that the tests of self-consistency provided 
by the agreement of the results of the two analyses are indeed non-trivial.

Below, the magnitude of the $D=6$ contributions will be estimated using the
vacuum saturation approximation (VSA) with very generous errors, and 
$D=8$ and $D=10$ contributions will be assumed negligible.{\footnote
{Existing fits for the effective $D=6$ and $8$ condensates $C_{6,8}$
for the $ud$ $V$ and $A$ correlators can be used to evaluate the 
$ud$ contributions to the FB $D=6+8$ differences. These results will be 
used as a very conservative bound for the corresponding FB combinations.}}
The fact that higher $D$ contributions vary more strongly with $s_0$ 
than do integrated lower dimension $D=2,\ 4$ contributions means that these 
assumptions can also be tested, provided a range of $s_0$ is employed in the 
analysis. As already emphasized above, such $s_0$-stability tests play 
an important role in all such sum-rule analyses.

For $T=V\pm A$, $\pi$ and $K$ pole contributions appear on both sides of 
Eq.~(\ref{imsrbasic}). The LECs of interest in these cases, in fact, appear 
only in the pole-subtracted parts, $\Delta\overline{\Pi}_{V\pm A}$, of the 
$\Delta\Pi_{V\pm A}$, and it is thus convenient to move all the pole 
contributions to the corresponding RHSs. With $w_{DK}(0)=\hat{w}(0)=1$, the 
three IMFESRs of interest, for $w(s)=w_{DK}(y)$ or $\hat{w}(y)$, then take 
the form
\begin{eqnarray}
\Delta\Pi_V(0) &&= {\frac{1}{2\pi i}}\,\oint_{\vert s\vert = s_0} ds\,
{\frac{w(s)}{s}}\, \left[\Delta\Pi_V (Q^2)\right]^{OPE}\, +\, 
\int_{th}^{s_0}ds\, {\frac{w(s)}{s}}\,\Delta{\rho}_V(s)\ ,\label{dkvsr}\\
\Delta\overline{\Pi}_{V\pm A}(0) &&= 
{\frac{1}{2\pi i}}\,\oint_{\vert s\vert = s_0} ds\,
{\frac{w(s)}{s}}\, \left[\Delta\Pi_{V\pm A} (Q^2)\right]^{OPE}\, +\, 
\int_{th}^{s_0}ds\, {\frac{w(s)}{s}}\,\Delta\overline{\rho}_{V\pm A}(s)
\nonumber\\
&&\ \ \ \ \ \pm \left[{\frac{f_K^2}{s_0}}\, f^w_{res}(y_K)\,
-\, {\frac{f_\pi^2}{s_0}}\, f^w_{res}(y_\pi )\right]\ ,
\label{vpmaimsralt}\end{eqnarray}
where, owing to the pole subtraction, $th$ is now the continuum
threshold $4m_\pi^2$, $y_\pi =m_\pi^2/s_0$, 
$y_K=m_K^2/s_0$, $f^{w_{DK}}_{res}(y)=4-y-y^2-y^3+y^4$ and
$f^{\hat{w}}_{res}(y)=6-6y+2y^2$. The normalization of the decay constants 
is such that $f_\pi \approx 92$ MeV. Note that it is the full
correlators $\Delta\Pi_{V\pm A}$, including the $\pi$/$K$
pole contributions, which occur in the first term of
the RHS of Eq.~(\ref{vpmaimsralt}). We discuss the inputs to the RHSs of 
Eqs.~(\ref{dkvsr}) and (\ref{vpmaimsralt}) in the next section.

We conclude this section with the NNLO low-energy representations of the
LHSs of Eqs.~(\ref{dkvsr}) and (\ref{vpmaimsralt}). The general structure 
of these representations is
$R^T(0) \, +\, \left[\Delta\Pi_T(0)\right]_{LEC}$, with
\begin{eqnarray}
&&\left[\Delta\Pi_T(0)\right]_{LEC}\, =\,
\sum_{k=5,9,10}c^T_k L^r_k\, + \, 32(m_K^2-m_\pi^2)\, 
\sum_{k=12,61,80} a^T_k C^r_k\ ,
\end{eqnarray}
and $R^T(0)$, the NLO LECs $L^r_k$ and the NNLO LECs $C^r_k$ all depending
on the chiral renormalization scale $\mu$.
The $R^T(0)$ combine all one- and two-loop contributions involving
only LO vertices, and are completely fixed by the pseudoscalar meson 
masses, decay constants and $\mu$. Their forms are rather lengthy
(especially for the $T=V\pm A$ cases) and hence not presented here. 
They can be reconstructed from the results of Sections 4, 
6 and Appendix B of Ref.~\cite{abt00}. NLO contributions proportional
to the $L_k^r$ cancel in the FB combinations considered here. 
The coefficients $c^T_{5,9,10}$ are generated by one-loop graphs with a single 
NLO vertex, and thus also $\mu$-dependent. 
The $a^T_k$ are, in contrast, $\mu$-independent at this order.

Fixing the chiral scale $\mu$ to the conventional scale choice 
$\mu_0 \equiv 0.77$ GeV, the 
explicit forms of the $\left[\Delta\Pi_T(0)\right]_{LEC}$ become
\begin{eqnarray}
&&\left[\Delta\Pi_V(0)\right]_{LEC}\, =\, -0.7218\, L_5^r\,
+\, 1.423\, L_9^r\, +\, 1.062\, L_{10}^r 
\, +\, 32(m_K^2-m_\pi^2)C_{61}^r\ ,\nonumber\\
&&\left[\Delta{\overline{\Pi}}_{V+A}(0)\right]_{LEC}\, =
\, -0.7218\, L_5^r\, +\, 1.423\, L_9^r
\, +\, 32(m_K^2-m_\pi^2)\left[ C_{12}^r+C_{61}^r
+C_{80}^r\right]\ ,\nonumber\\
&&\left[\Delta{\overline{\Pi}}_{V-A}(0)\right]_{LEC}\, =
\, -0.7218\, L_5^r\, +\, 1.423\, L_9^r\, +\, 2.125\,
L_{10}^r\nonumber\\
&&\qquad\qquad\qquad\qquad\qquad -\, 32(m_K^2-m_\pi^2)\left[ C_{12}^r-C_{61}^r
+C_{80}^r\right]\ ,
\label{nlonnlolectermcontributions}\end{eqnarray}
where the renormalized LECs are all understood to be evaluated at
$\mu=\mu_0$, and $m_\pi$ and $m_K$ have been taken to be the charged 
pion mass and, for definiteness, the average of the charged and neutral kaon 
masses. (Taking instead $m_K$ to be the charged $K$ mass 
has no impact on our final results.) The corresponding values for the 
LEC-independent $R_T$ contributions are
\begin{eqnarray}
&&R_V(\mu_0 )\, =\, 0.00775\, ,\nonumber\\
&&R_{V+A}(\mu_0)\, =\, 0.00880\, , \nonumber\\
&&R_{V-A}(\mu_0)\, =\, 0.00670\label{nonlecudmusrzerovalues}\ .
\end{eqnarray}

\section{\label{input}Inputs to the $V$, $V+A$ and $V-A$ IMFESRs}
\subsection{\label{nlolecinput}Meson masses, decay constants and NLO LEC
inputs}
PDG 2012 values~\cite{pdg2012} are used for $f_\pi$, $m_\pi$, $m_K$ and 
$m_\eta$ (the latter is required in evaluating some of the NNLO 
contributions). Explicitly, $m_\pi=139.57$ MeV, $m_\eta =547.85$ MeV 
and $m_K=495.65$ MeV, the latter being the average of the charged and 
neutral masses. For the normalization used here, $f_\pi = 92.21(14)$ 
MeV~\cite{pdg2012}. $f_K$ is obtained by combining this value  
with the current FLAG assessment of $n_f=2+1$ lattice results, 
$f_K/f_\pi = 1.193(5)$~\cite{flag}. 

The NNLO representation of $\Delta\overline{\Pi}_{V+A}(0)$ does not
involve $L_{10}^r$, so the only NLO LECs required as input to the 
determination of the corresponding NNLO LEC combination 
$C_{12}^r+C_{61}^r+C_{80}^r$ are $L_5^r$ and $L_9^r$. The $\Delta\Pi_V(0)$ 
and $\Delta\overline{\Pi}_{V-A}(0)$ IMFESRs, in contrast, also 
require input on $L_{10}^r$.

For $L_5^r$ and $L_9^r$, we employ the values
$L_5^r(\mu_0)=0.00058(13)$ and $L_9^r(\mu_0)=0.00593(43)$. 
The former is the result of the recommended {\it All} fit from the most
recent NNLO analysis of the NLO LECs $L^r_{1-8}$, given in Table 5 of 
Ref.~\cite{bj11}. The latter was obtained in an NNLO analysis of the 
$\pi$ and $K$ electromagnetic form factors~\cite{bt02}.
The contributions proportional to $L_5^r$ in the three IMFESRs under
consideration are numerically rather small, and the resulting
contributions to the errors on the corresponding NNLO LEC combinations 
negligible in comparison to the contributions from other sources.

The situation with $L_{10}^r$ is somewhat more complicated since the NNLO 
determination of $L_{10}^r$ is not independent of the NNLO LECs appearing 
in the $\Delta\Pi_V$ and $\Delta\overline{\Pi}_{V-A}$ IMFESRs. The standard 
route to an experimental determination of $L_{10}^r$ has been through a 
dispersive or IMFESR determination of the value of the $\pi$-pole-subtracted 
light-quark V-A correlator, $\overline{\Pi}_{ud;V-A}(Q^2)$, 
at $Q^2=0$. An early IMFESR analysis, employing ChPT to NLO 
may be found in Ref.~\cite{dghs98}. Two NLO determinations using lattice 
data for $\Pi_{ud;V-A}(Q^2)$ also exist~\cite{jlqcdvmanlo,rbcukqcdvmanlo}. 
A very precise determination, 
\begin{equation}
\overline{\Pi}_{ud;V-A}(0)\, =\, 0.0516(7)\ ,
\label{pivmaudzero}\end{equation}
has been obtained in Ref.~\cite{bgjmp13} using the results of 
Refs.~\cite{dv71,dv72} in combination with the updated version of the OPAL 
non-strange spectral distributions~\cite{opalud99} reported in 
Ref.~\cite{dv72}. A similar result for $\Pi_{ud;V-A}(0)$ has been obtained in
Ref.~\cite{gapp08} using the non-strange ALEPH, rather than OPAL, 
data~\cite{alephud98}. The error on the result of Ref.~\cite{gapp08} is,
unfortunately, not reliable, owing to an error in the publicly posted
covariance matrices for the version of the ALEPH data used in that
analysis~\cite{dv7tau10}. 

It is now known that 
the NLO approximation provides a very poor representation 
of the low-$Q^2$ dependence of $\overline{\Pi}_{ud;V-A}(Q^2)$~\cite{bgjmp13}.
This result, which is not unexpected in view of a similar observation
about the NLO representation of the $ud$ V correlator~\cite{ab06}, 
clearly calls into question the results for $L_{10}^r$ 
obtained from NLO analyses.

The NNLO representation of $\overline{\Pi}_{ud;V-A}(0)$ required to extend 
the NLO analyses to NNLO has the form~\cite{abt00} 
\begin{equation}
\overline{\Pi}_{ud;V-A}(0) \, =\, 
{\cal R}_{ud;V-A} \, +\, \hat{c}_{9} L^r_9\, +\, 
\hat{c}_{10}L^r_{10} \, +\, {\cal C}^r_0\, +\, {\cal C}^r_1\, , 
\label{udvmannlo}\end{equation}
where ${\cal R}_{ud;V-A}$ is the sum of one- and two-loop contributions
involving only LO vertices, 
\begin{eqnarray}
&&\hat{c}_9\, =\, 16\left(2 \mu_\pi +\mu_K\right)\, ,\nonumber\\
&&\hat{c}_{10}\, =\, -8\left(1-8 \mu_\pi -4\mu_K\right)\ ,
\label{l9l10udvmacoeffs}\end{eqnarray}
with $\mu_P = {\frac{m_P^2}{32\pi^2f_\pi^2}}\, \log\left({\frac{m_P^2}
{\mu^2}}\right)$ the usual chiral logarithm, and 
\begin{eqnarray}
{\cal C}^r_0\, &&=\, 32m_\pi^2\left[
C_{12}^r-C_{61}^r+C_{80}^r\right]\, ,\nonumber\\
{\cal C}^r_1\, &&=\, 32\left( m_\pi^2+2m_K^2\right)\, 
\left[ C_{13}^r-C_{62}^r+C_{81}^r\right]\ . 
\label{nnlolecudvma}\end{eqnarray}
${\cal R}_{ud;V-A}$, $\hat{c}_9$ and $\hat{c}_{10}$ are all fixed by the
chiral scale $\mu$ and pseudoscalar masses and decay constants. The 
NNLO LECs appearing in ${\cal C}^r_0$ are LO in $1/N_c$, those in 
${\cal C}^r_1$ $1/N_C$-suppressed. Note that ${\cal C}^r_0$ involves precisely 
the combination of NNLO LECs appearing in $\Delta\overline{\Pi}_{V-A}(0)$. For 
$\mu=\mu_0$, the results of Ref.~\cite{bgjmp13} for 
$\overline{\Pi}_{ud;V-A}(0)$ and Ref.~\cite{bt02} for $L_9^r(\mu_0 )$
yield the very precise constraint 
\begin{equation}
L_{10}^r(\mu_0 )\, =\, -0.004098(59)_{exp}(74)_{L_9^r}\, +\,
0.0822\left({\cal C}^r_0+{\cal C}^r_1\right)
\label{udvmaq20constraint}\end{equation}
on $L_{10}^r(\mu_0 )$, ${\cal C}^r_0(\mu_0 )$ and 
${\cal C}^r_1(\mu_0 )$.{\footnote{The slight difference between the
result given in Eq.~(\ref{udvmaq20constraint}) and that,
$L_{10}^r\, =\, -0.004113(89)_{exp}(74)_{L_9^r}$, quoted in Eq.~(4.9) of 
Ref.~\cite{bgjmp13}, results from the inadvertent use in Ref.~\cite{bgjmp13} 
of the less precise determination of the quantity $L_{10}^{\rm eff}$, given
by Eq.~(4.1) of that reference, in place of the most precise determination,
Eq.~(4.2b). Switching instead to the most precise determination, Eq.~(4.2b),
leads to the result quoted in Eq.~(\ref{udvmaq20constraint}).}}
Information on ${\cal C}^r_0(\mu_0 )$ and ${\cal C}^r_1(\mu_0 )$ is, however, 
required to turn this into a determination of $L_{10}^r(\mu_0 )$. The 
differing dependences of $\hat{c}_9$, $\hat{c}_{10}$, ${\cal C}^r_0$ and 
${\cal C}^r_1$ on the meson masses makes it natural to approach this problem 
using the lattice, where the pseudoscalar meson masses can be varied
by varying the input quark masses. 

Such an analysis has been carried out in Ref.~\cite{boylelatt13vma}. 
The first stage of this analysis uses lattice and continuum data for 
$\overline{\Pi}_{ud;V-A}(Q^2)$ in combination with the constraint, 
Eq.~(\ref{udvmaq20constraint}), obtained from the already published 
result for $\overline{\Pi}_{ud;V-A}(0)$, Eq.~(\ref{pivmaudzero}). For 
low Euclidean $Q^2$, the errors on the lattice data for 
$\overline{\Pi}_{ud;V-A}(Q^2)$ are currently larger than those on the 
continuum version. The consequence is that, while use of the lattice 
data in combination with the $\overline{\Pi}_{ud;V-A}(0)$ constraint,  
Eq.~(\ref{udvmaq20constraint}), {\it does} allow all three of $L_{10}^r$, 
${\cal C}_0^r$ and ${\cal C}_1^r$ to be determined, the errors that result 
from this first stage analysis are at the $\sim 25\%$, $\sim 100\%$ and 
$\sim 80\%$ levels for $L_{10}^r$, ${\cal C}_0^r$ and ${\cal C}_1^r$, 
respectively. The $\Delta\overline{\Pi}_{V-A}$ IMFESR, which involves a 
distinct combination of two of these three quantities, $L_{10}^r$ and 
${\cal C}^r_0$, provides an additional constraint, and allows an extended
(second stage) version of the analysis of Ref.~\cite{boylelatt13vma}
to be carried out. The extended analysis, which employs our results below 
for the $\Delta\overline{\Pi}_{V-A}$ IMFESR constraint as input, produces 
results (quoted in Eqs.~(\ref{l10rcombined}), (\ref{scriptC0combined})
and (\ref{scriptC1combined}) below) with significantly reduced errors.

In presenting the results of the IMFESR analyses below, we will thus first 
quote the result for $C_{12}^r+C_{61}^r+C_{80}^r$ from the
$\Delta\overline{\Pi}_{V+A}$ IMFSER, which is independent of the
treatment of $L_{10}^r$, and then quote the result for the constraint on 
$L_{10}^r$ and ${\cal C}^r_0$ arising from the $\Delta\overline{\Pi}_{V-A}$ 
IMFESR. The $\Delta\overline{\Pi}_{V+A}$ and $\Delta\overline{\Pi}_{V-A}$ 
IMFESRs can, of course, also be combined to obtain the related (but not 
independent) $\Delta\Pi_V$ and $\Delta\overline{\Pi}_A$ IMFESRs. The former 
constrains the combination of $L_{10}^r$ and $C_{61}^r$ noted above, the 
latter an analogous combination of $L_{10}^r$ and $C_{12}^r+C_{80}^r$. To 
go further, and turn these constraints into explicit determinations of the 
corresponding NNLO LEC combinations, requires input on $L_{10}^r$. We will 
employ for $L_{10}^r$ the result of the second stage combined 
lattice-continuum analysis of Ref.~\cite{boylelatt13vma}.
This analysis, incorporates the $w_{DK}(y)$ version of the 
$\Delta\overline{\Pi}_{V-A}$ IMFESR constraint, in addition to the
$\overline{\Pi}_{ud;V-A}(0)$ constraint, Eq.~(\ref{udvmaq20constraint}),
and the constraints generated by data from four different lattice
ensembles. The resulting error for $L_{10}^r$ is
dominated by lattice errors, and hence independent of those in 
the present analysis. With this input for $L_{10}^r$, $C_{61}^r$ 
follows from the $\Delta\Pi_V$ IMFESR constraint and $C_{12}^r+C_{80}^r$ 
from the $\Delta\overline{\Pi}_A$ IMFESR constraint. External 
input on $C_{12}^r$ then allows us to also fix $C_{80}^r$. 

\subsection{\label{OPEinput}OPE input}
The correlator combinations entering the IMFESRs under consideration are all
flavor-breaking and thus have vanishing $D=0$ OPE series. We include $D=2$ 
and $4$ contributions for all channels, treat $D=6$ and $8$ contributions as 
discussed below, and assume that $D=10$ and higher contributions can be 
neglected for IMFESRs based on either $w_{DK}(y)$ or $\hat{w}(y)$.
Integrated duality violations will also be neglected. Since integrated OPE 
contributions of $D=2k$ scale, up to logarithms, as $1/s_0^k$ 
(see Eq.~(\ref{higerdopeint})), and integrated duality violations
typically produce contributions with oscillatory $s_0$-dependence, 
these assumptions can be tested by studying the IMFESRs
(\ref{dkvsr}) and (\ref{vpmaimsralt}) over a range of $s_0$ and ensuring 
that the resulting $Q^2=0$ correlator values are independent of $s_0$, 
as well as of the choice of weight, as they should be.

The $D=2$ OPE series for the flavor $ij=ud,us$, $J=0+1$, V and A correlators
are known to four loops. The explicit expressions to three loops, including
light-quark mass corrections, may be found in Ref.~\cite{ck98}, and the
$O(m_s^2)$ terms in the four-loop contributions in Ref.~\cite{bck05}. 
Expressions for the corresponding $D=4$ and $6$ contributions may be
found in Refs.~\cite{deq4refs,bnp}.

Omitting, for presentational simplicity, corrections suppressed by one or 
more powers of $m_{u,d}/m_s$, these results imply, for $D=2$,
\begin{eqnarray}
&&\left[\Delta\Pi_V(Q^2)\right]^{OPE}_{D=2}\, =\, {\frac{3}{4\pi^2}}
{\frac{m^2_s(Q^2)}{Q^2}}\, \left[ 1\, +\, {\frac{7}{3}}\bar{a}
\, +\, 19.9332\bar{a}^2\, +\, 208.746\bar{a}^3\, +\, \cdots\right]\, ,
\nonumber\\
&&\left[\Delta\Pi_{V+A}(Q^2)\right]^{OPE}_{D=2}\, =\, {\frac{3}{2\pi^2}}
{\frac{m^2_s(Q^2)}{Q^2}}\, \left[ 1\, +\, {\frac{7}{3}}\bar{a}
\, +\, 19.9332\bar{a}^2\, +\, 208.746\bar{a}^3\, +\, \cdots\right]\, ,
\nonumber\\
&&\left[\Delta\Pi_{V-A}(Q^2)\right]^{OPE}_{D=2}\, =\, {\frac{3}{2\pi^2}}
{\frac{m_u(Q^2) m_s(Q^2)}{Q^2}}\, \left[ {\frac{2}{3}}\bar{a}
\, +\, 8.7668\bar{a}^2\, +\, \cdots\right]\, ,
\label{d2opeforms}\end{eqnarray}
where $\bar{a}\equiv \alpha_s(Q^2)/ \pi$, with $\alpha_s(Q^2)$
the $\overline{MS}$ running coupling, and $m_u(Q^2)$ and $m_s(Q^2)$ are 
the $\overline{MS}$ running $u$ and $s$ quark masses. 

For $D=4$, one has, omitting numerically negligible contributions 
fourth order in the quark masses and terms suppressed by $m_{u,d}/m_s$,
\begin{eqnarray}
&&\left[\Delta\Pi_V(Q^2)\right]^{OPE}_{D=4}\, =\, -{\frac{1}{Q^4}}\,
\left({\frac{m_s}{\hat{m}}}\right)\, \langle \hat{m}\bar{u}u\rangle
\, \left[ r_c\, +\, \bar{a}\left({\frac{4}{3}}-r_c\right)
\, +\, \bar{a}^2\left({\frac{59}{6}}-{\frac{13}{3}}r_c\right)\right]\, ,
\nonumber\\
&&\left[ \Delta\Pi_{V+A}(Q^2)\right]^{OPE}_{D=4}\, =\, -{\frac{2}{Q^4}}\,
\left({\frac{m_s}{\hat{m}}}\right)\, r_c \, \langle \hat{m}\bar{u}u\rangle
\, \left[ 1-\bar{a}-{\frac{13}{3}}\bar{a}^2\right]\, ,\nonumber\\
&&\left[\Delta\Pi_{V-A}(Q^2)\right]^{OPE}_{D=4}\, =\, -{\frac{1}{Q^4}}\,
\left({\frac{m_s}{\hat{m}}}\right)\, \langle \hat{m}\bar{u}u\rangle
\, \left[ {\frac{8}{3}}\bar{a}+{\frac{59}{3}}\bar{a}^2\right]\, ,
\label{d4opeforms}\end{eqnarray}
where $\hat{m}=(m_u+m_d)/2$ and $r_c=\langle \bar{s}s\rangle / \langle\bar{u}u
\rangle$. 

$D=6$ contributions are expected to be dominated by contributions from 
four-quark condensates. These condensates are not known experimentally for 
the flavor $us$ correlators, but can be roughly estimated using the VSA.
In this approximation one has~\cite{bnp}
\begin{eqnarray}
&&\left[ \Pi_{ij;V/A}^{(0+1)}(Q^2)\right]^{OPE}_{D=6,VSA}\, =\, 
{\frac{32\pi}{81}}\, {\frac{\alpha_s}{Q^6}}\, \left[
\mp 9\langle \bar{q}_iq_i\rangle\langle \bar{q}_jq_j\rangle
+ \langle \bar{q}_iq_i\rangle^2 +\langle \bar{q}_jq_j\rangle^2\right]\ ,
\label{d6opeudus0p1va}\end{eqnarray}
from which the VSA approximations to $\left[\Delta\Pi_V\right]^{OPE}_{D=6}$
and $\left[\Delta\Pi_{V\pm A}\right]^{OPE}_{D=6}$ are easily obtained.
With these estimates, one finds that $D=6$ contributions are numerically 
very small, particularly so for the $V$ and $V+A$  IMFESRs, where they
could be safely neglected even if the VSA were to be in error by
an order of magnitude.{\footnote{In fact, the VSA turns out
to yield central values for the $D=6$ contributions much smaller
than the corresponding estimated errors. To the number of digits
quoted below, our final results are, in fact, unchanged if we shift
from our VSA estimates to zero for the $D=6$ contributions.}}

The VSA estimates for the $D=6$ contributions to the FB IMFESRs, being
proportional to the FB factor $r_c-1$, display quite strong 
cancellations. Some care must thus be exercised in assigning errors 
to these estimates. Here it is possible to take advantage of recent 
results for the effective $D=6$ and $D=8$ condensates appearing in 
the OPE representations of $\Pi_{ud;V}$, $\Pi_{ud;A}$ and $\Pi_{ud;V-A}$, 
obtained in the course of the analyses described in Refs.~\cite{dv72,bgjmp13}. 
These allow a determination of the sum of $D=6$ and $8$ contributions to the 
$ud$ parts of the relevant IMFESR OPE integrals. Although the corresponding 
flavor $us$ contributions are not known, some degree of cancellation will
certainly be present in the FB $ud-us$ differences. The flavor $ud$
$D=6+8$ OPE sums can thus be used to provide a very conservative
estimate of the uncertainties on the central FB $ud-us$ $D=6+8$ OPE 
contributions described above. 

The inputs required to evaluate the $D=2,\, 4$ OPE contributions are as 
follows. For the running coupling and masses we employ the exact solutions 
generated using the four-loop-truncated $\beta$ and $\gamma$ 
functions~\cite{4loopbetagamma}. The initial condition for $\alpha_s$ 
is taken to be $\alpha_s^{n_f=3}(m_\tau^2)=0.3181(57)$, obtained from 
the $n_f=5$ PDG 2012 assessment $\alpha_s(m_Z^2)^{n_f=5}=0.1184(7)$ 
via standard four-loop running and three-loop matching at the flavor 
thresholds~\cite{cks97}. For the initial conditions for the running 
masses we take the results for $m_{u,d,s}(2\ {\rm GeV})$ contained in 
the latest FLAG assessment~\cite{flag}. The GMOR relation 
$\hat{m}\langle \bar{u}u\rangle\, =\, -{\frac{1}{2}}m_\pi^2f_\pi^2$ is
used for the light-quark condensate. For the ratio of strange to light 
condensates, the recent lattice result, $r_c = 1.08(16)$~\cite{davies12},
is in good agreement with the value $r_c=1.1(3)$ obtained by updating the 
sum rule result of Ref.~\cite{jl02} for modern $n_f=2+1$ values of the
ratio $f_{B_s}/f_B$. To be conservative, we will take the larger of the
two errors. 

In the case of the $V$ and $V+A$ IMFESRs, the largest source of uncertainty 
in the OPE contribution turns out to lie in the treatment of the integrated 
$D=2$ series. Since $\alpha_s(m_\tau^2)/\pi \simeq 0.1$, one sees from 
Eqs.~(\ref{d2opeforms}) that, in these cases, the convergence of the
known terms in the $D=2$ series is marginal at best: at the spacelike
point on the contour, the four-loop ($O(\bar{a}^3)$) $D=2$ term in fact 
exceeds the three-loop ($O(\bar{a}^2)$) one for all $s_0$ accessible using 
$\tau$-decay data. The rather problematic convergence behavior of the $D=2$ series 
manifests itself not only in a similarly problematic behavior for the 
integrated $D=2$ series, but also in a large difference, increasing with 
truncation order, between the results of evaluations of the integrated 
truncated series obtained using the FOPT (fixed-order perturbation theory)
and CIPT (contour-improved perturbation theory) prescriptions. The 
two prescriptions differ only by contributions of order higher than
the truncation order, the former involving the truncation of the
integrated series at fixed order in $\alpha_s(s_0)$, the latter 
the summation of logarithms point-by-point along the contour
via the local scale choice $\mu^2 =Q^2$ and truncation at the same fixed
order in $\alpha_s(Q^2)$ for all such $Q^2$. 

The problematic convergence behavior and increase in the FOPT-CIPT difference 
with increasing truncation order both suggest the $D=2$ series may already 
have begun to display its asymptotic character at three- or four-loop order,
complicating an assessment of the error to be assigned to the integrated
truncated series. This issue has been raised previously in the context of the 
determination of $\vert V_{us}\vert$ from FB hadronic $\tau$ decay 
sum rules~\cite{kmvustau}. 

Fortunately, the lattice provides a means of investigating the reliability
of various treatments of the $D=2$ OPE series. In Refs.~\cite{boyled2convlatt}
lattice data for the FB $V+A$ combination was shown to favor the
fixed-scale over local-scale treatment of the $D=2$ series, hence FOPT over 
CIPT for the IMFESR integrals. Moreover, with the three-loop-truncated,
fixed-scale version of the $D=2$ series, the OPE was found to provide a 
good representation of the lattice data for $Q^2$ in the range
from $m_\tau^2$ down to $\sim 2\ {\rm GeV}^2$~\cite{boyled2convlatt}.
In view of these results, the integrated $D=2$ OPE contribution 
has been evaluated using the FOPT prescription truncated at three loops.
The associated error is taken to be the quadrature sum of (i) the
three-loop FOPT-CIPT difference, (ii) the magnitude of the last (three-loop) 
term retained in the integrated FOPT series, (iii) the error 
associated with the uncertainty in the input $m_s(2\ {\rm GeV})$, and (iv) the 
error associated with the uncertainty in the input $\alpha_s(m_\tau^2)$. 
The resulting error is dominated by the FOPT-CIPT difference for $w_{DK}$,
while both the FOPT-CIPT difference and last-term-retained contributions 
are important for $\hat{w}$. Based on the lattice results, this approach 
should, in fact, yield a rather conservative assessment of the $D=2$ error.

Further details of our assessments of the errors on the various OPE 
contributions may be found in the Appendix.

\subsection{\label{spectralinput}Flavor $ud$ and $us$ spectral input}
The weighted spectral integrals needed to complete the evaluations of 
the RHSs of the IMFESRs Eqs.~(\ref{vpmaimsralt}) are  
\begin{eqnarray}
&&\int_{th}^{s_0}ds\, {\frac{w(s)}{s}}\, \rho_{ud,us;V}^{(0+1)}(s)
\quad {\rm and} \quad\int_{th}^{s_0}ds\, {\frac{w(s)}{s}}\, 
\overline{\rho}_{ud,us;A}^{(0+1)}(s)\ , 
\label{neededspecints}\end{eqnarray}
with $w(s)=w_{DK}(y)$ or $\hat{w}(y)$, and the range 
$2.15\ {\rm GeV}^2<s_0<m_\tau^2$ (the lower bound
reflecting the binning of the ALEPH data) employed to carry out
the $s_0$-stability (self-consistency) tests noted above.

An update of the OPAL results for $\rho_{ud;V}^{(0+1)}(s)$ and
$\overline{\rho}_{ud;A}^{(0+1)}(s)$~\cite{opalud99}, reflecting changes to
the exclusive-mode branching fractions since the original OPAL
publication, was performed in Ref.~\cite{dv72}. This update
employed non-strange branching fractions from an HFAG fit incorporating
Standard Model expectations based on $\pi_{\mu 2}$ and $K_{\mu 2}$ decay
widths and then-current strange branching fractions from the
same fit. Since then, Belle has produced a new result for 
$B[\tau^-\rightarrow K_S\pi^-\pi^0\nu_\tau ]$~\cite{bellekspipi}
which shifts slightly the previous world average for this mode. 
To restore the sum over all branching fractions to one after this
shift, and in the absence of an update of the previously used HFAG fit 
which takes this shift into account, a common global $0.99971$ rescaling 
has been performed on the $ud$ $V$ and $A$ distributions of Ref.~\cite{dv72}.
Being so close to one, this rescaling, not surprisingly, has negligible 
effect.

The $V/A$ separation for the non-strange modes was performed by
OPAL using G-parity. The main uncertainty in this separation 
results from $K\bar{K}\pi$ contributions, for which G-parity
cannot be used. A conservative, fully anticorrelated $50\pm 50\%$ 
$V/A$ breakdown was assumed. While the $K\bar{K}\pi$ $V/A$ separation
uncertainty can, in principle, be significantly reduced through
angular analyses~\cite{km92} of the much higher statistics B-factory data on 
these modes, such an improvement is irrelevant for our purposes
since this uncertainty plays a negligible role in the present analysis.

The differential decay distribution $dR_{us;V+A}/ds$ has been measured, 
and its exclusive mode contributions made available, by the ALEPH 
collaboration~\cite{alephus99}. Much higher statistics B-factory results 
now exist for the relative (unit-normalized) distributions of the 
$K^-\pi^0$~\cite{babarkmpi0}, $K_S\pi^-$~\cite{bellekspi}, 
$K^-\pi^+\pi^-$~\cite{iannugentthesis,babarkpipiallchg} and 
$K_S\pi^-\pi^0$~\cite{bellekspipi} exclusive modes, the latter
in preliminary form only. 
We employ the B-factory results for these four modes, using current 
branching fraction values to fix the overall normalizations.{\footnote{The 
version of the Belle $K_S\pi^-\pi^0$ results used here is preliminary,
having been read off from Fig. 2 of the report, Ref.~\cite{bellekspipi}, 
prepared by the Belle collaboration for the Tau 2012 proceedings. The 
errors take into account the uncertainty in the reported branching 
fraction in addition to those shown in the figure. While this should 
(and will) be updated once the final version of the Belle analysis is 
released, the $K_S\pi^-\pi^0$ uncertainties errors play a negligible 
role in the $V$ and $V-A$ analyses (where they are swamped by the much larger 
$V/A$ separation uncertainties) and a very small role in the $V+A$ analysis 
(where $K\pi$ error contributions are dominant). The preliminary
nature of the Belle data should thus have no relevant impact on the
present analysis.}}
For all other modes the ALEPH results, rescaled to current branching 
fraction values, are used. The $J=0$ subtraction of the $dR_{us;V+A}/ds$ 
distribution, required to extract the $J=0+1$ component contribution thereof, 
and hence the combination $\overline{\rho}_{us;V+A}^{(0+1)}(s)$ is, as noted
above, performed using the results of Refs.~\cite{jopss,kmsps}. For 
$\vert V_{us}\vert$ (needed to convert from $dR/ds$ to the spectral function) 
we employ the value, $0.2255(10)$, implied by three-family unitarity and the 
Hardy-Towner determination $\vert V_{ud}\vert = 0.97425(22)$~\cite{htvud}.

The $V/A$ separation of the $us$, $V+A$ distribution is more complicated than 
in the analogous $ud$ case. While the $K$ pole contribution is pure $A$,
and the $K\pi$ distribution pure $V$, chirally unsuppressed $V$ and $A$ 
contributions are both present for all the higher multiplicity 
$K\, n\pi$ ($n\ge 2$) modes. For $K\pi\pi$, the $V/A$ separation could be 
performed, up to small chirally suppressed corrections, by a relatively 
simple angular analysis~\cite{km92}, but this has yet to be done. Fortunately,
for phase space reasons, the $K\pi\pi$ and higher multiplicity
strange mode distributions lie at relatively high $s$, increasingly
so with increasing multiplicity. Their contributions to the IMFESR spectral 
integrals are thus strongly suppressed by the combination of the $1/s$ factor 
in the overall weight, $w(s)/s$, and the triple-zeros of $w_{DK}(y)$ and 
$\hat{w}(y)$ at $s=s_0$. The suppression of such high-$s$ contributions, 
of course, grows stronger as $s_0$ is decreased. The strong high-$s$ 
suppression is also welcome in view of the low statistics, and consequent 
large errors, for the high-$s$ part of the ALEPH $us$ distribution and
the fact that the $s$-dependences of the ALEPH $K\, 3\pi$, $K\eta$, 
$K\, 4\pi$ and $K\, 5\pi$ distributions were fixed from Monte Carlo
rather than by direct measurement. The high-$s$ suppression is, in fact, 
strong enough to allow the analyses to proceed with a $50\pm 50\%$ (fully 
anticorrelated) $V/A$ breakdown assigned to contributions from all modes
other than $K$ and $K\pi$. 

As noted in Refs.~\cite{alephus99,dk00}, however,
the $K\pi\pi$ distributions contain contributions from the axial $K_1(1270)$,
the axial $K_1(1400)$ and the vector $K^*(1410)$ resonances. While the latter 
two cannot be disentangled without an angular analysis, the former lies in 
a distinct part of the spectrum, and can be unambiguously assigned to the 
$A$ channel. This observation allows an improvement to be made on the $V/A$ 
separation for the $K\pi\pi$ modes. D\"urr and Kambor~\cite{dk00}, following
ALEPH~\cite{alephus99}, modelled the $K\pi\pi$ distribution as a sum of 
two resonant contributions, one from the $K_1(1270)$ and one from a single 
effective $1400$ region resonance with mass and width equal to the average
of the corresponding $K_1(1400)$ and $K^*(1410)$ parameters. The resulting 
$1400$ region contribution was then assigned $50\pm 50\%$ each to the $V$ 
and $A$ channels. For the $V$ channel considered by D\"urr and Kambor, the 
resulting ALEPH-based $K\pi\pi$ IMFESR contribution was found to be only 
$\sim 5\%$ ($\sim 7\%$) of the corresponding $K\pi$ one at 
$s_0\sim 2\ {\rm GeV}^2$ ($s_0\sim m_\tau^2$). 

\begin{figure}[H]
\caption{\label{kpipidistribs}The $K^-\pi^+\pi^-$ and $\bar{K}^0\pi^-\pi^0$
contributions to $\rho_{us;V+A}^{(0+1)}(s)$ implied
by the BaBar $K^-\pi^+\pi^-$~\cite{babarkpipiallchg} and Belle
$\bar{K}^0\pi^-\pi^0$~\cite{bellekspipi} results presented at
Tau 2012.}
\centering
{\rotatebox{270}{\mbox{
\includegraphics[width=3.25in]
{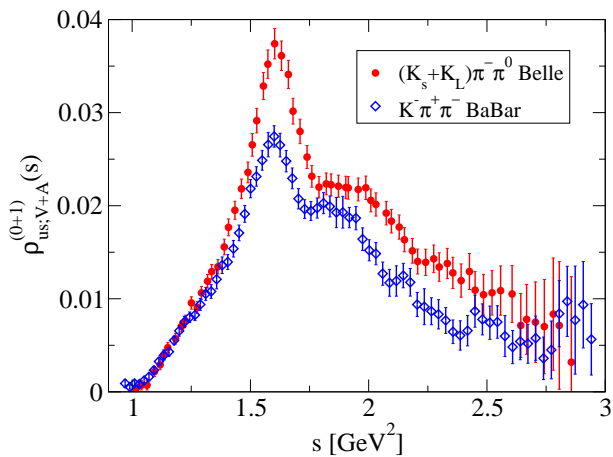}
}}}
\end{figure}

This approximate separation of $V$ and $A$ contributions to the $K\pi\pi$ 
spectral distribution can be carried out even more convincingly with the 
much higher precision BaBar $K^-\pi^+\pi^-$~\cite{babarkpipiallchg} and Belle
$K_S\pi^-\pi^0$~\cite{bellekspipi} data, both presented at Tau 2012. 
Fig.~\ref{kpipidistribs} shows the $us$ spectral function contributions 
produced by these data sets. The $K_1(1270)$ peak is clearly visible
for both modes. Performing the ALEPH/D\"urr-Kambor analysis, one finds $V/A$ 
breakdowns of $\sim$ $20\pm 20\%$ $V$/$80\pm 20\%$ $A$ for the 
$w_{DK}$- and $\hat{w}$-weighted IMFESR integral contributions from the 
combination of these two modes, the precise value varying by a few percent 
with variations in the choice of weight, the input 
effective $1400$ width and fit window employed, and by $\sim 2\%$ over 
the range of $s_0$ considered in this analysis. While the reduction
from $\pm 50\%$ to $\pm 20\%$, accomplished by taking into account the 
presence of the $K_1(1270)$ contributions, represents a significant 
improvement in the $V/A$ separation uncertainty for the $K\pi\pi$
component of the $us$ spectral integrals, one should bear in mind
that the $K\pi\pi$ contribution is much smaller than the $K$ and/or $K\pi$ 
ones, making the impact of this improvement on the errors in the total 
$us$ spectral integrals much more modest.

In the absence of high-statistics B-factory results for the distributions of
the much smaller $K^-\pi^0\pi^0$ mode, and all higher-multiplicity modes, we 
take the maximally conservative approach and assume a fully anti-correlated
$50\pm 50\%$ $V$/$50\pm 50\%$ $A$ breakdown for the corresponding spectral 
integral contributions. Because of the anticorrelation, both for these modes, 
and in the separation of the $1400$ region $K\pi\pi$ contributions, the total 
$us$ spectral integral error is magnified for the $V-A$ difference. For the
$V$, $A$ and $V-A$ channels, where the $V/A$ separation uncertainty plays a 
role, the suppression of contributions from the high-$s$ region produced by 
the triple zeros of $w_{DK}(y)$ and $\hat{w}(y)$ at $s=s_0$ and the $1/s$ 
factor in the full weight $w(s)/s$ is especially important. The $V/A$ 
separation uncertainty is, of course, absent for the V+A combination.

\section{\label{results}Results}
The RHSs of the IMFESRs of Eqs.~(\ref{dkvsr}) and (\ref{vpmaimsralt}) are 
evaluated using the input specified in the previous section. Included in 
this input is the choice of the three-loop-truncated FOPT prescription for 
evaluating the $D=2$ series. Since this choice was predicated on an agreement 
of the corresponding OPE representation and lattice data for Euclidean $Q^2$ 
extending from $m_\tau^2$ down to, but not below, $\sim 2\ {\rm GeV}^2$, we 
restrict our attention to $s_0$ lying safely in this interval. With the 
ALEPH $us$ data binning, this corresponds to 
$2.15\ {\rm GeV}^2\le s_0\le m_\tau^2$. For $s_0$ in this
range, experience with sum rules involving weights with a double zero at 
$s=s_0$ suggests integrated duality violations should also be  
negligible~\cite{dv71,dv72,kmalphastau}. OPE contributions are
very small for the $w_{DK}(y)$ version of the $\Delta\overline{\Pi}_{V-A}$ 
IMFESR, but less so for the $\hat{w}(y)$ version, where enhanced $D=4$ 
contributions reach up to $\sim 8\%$ of the RHS in the $s_0$ window
employed. OPE contributions are numerically relevant for both versions of
the $\Delta\Pi_V$ and $\Delta\overline{\Pi}_{V+A}$ IMFESRs, reaching 
$6\%$ and $8\%$, respectively, of the RHSs for the $w_{DK}(y)$ 
case, and $16\%$ and $19\%$, respectively, of the RHSs
for the $\hat{w}(y)$ case.

\begin{figure}[H]
  \begin{minipage}[t]{0.46\linewidth}
{\rotatebox{270}{\mbox{
\includegraphics[width=0.9\textwidth]{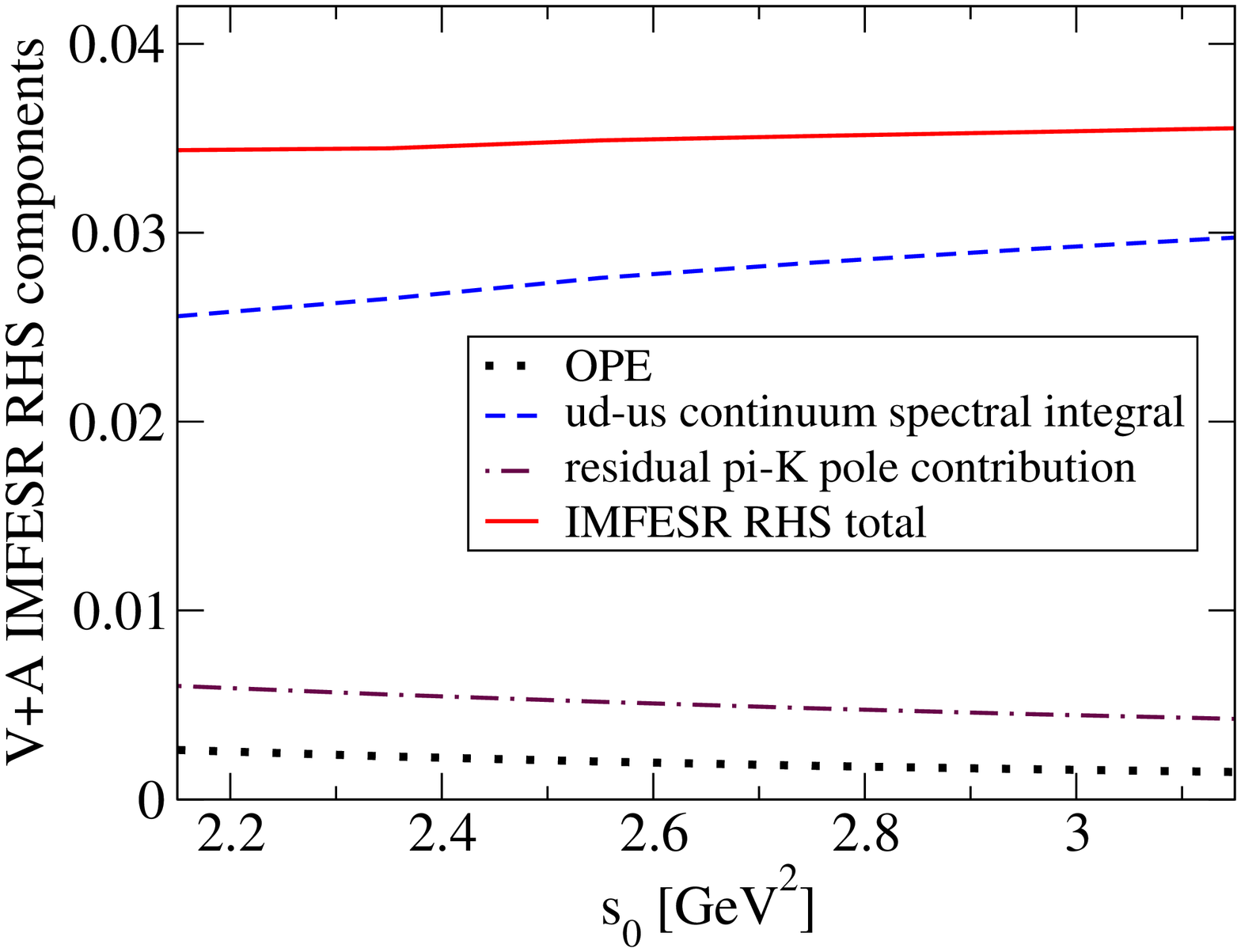}
}}}
  \caption{RHS contributions to the $w_{DK}(y)$
$\Delta\overline{\Pi}_{V+A}$ IMFESR, Eq.~(\ref{vpmaimsralt}), 
as a function of $s_0$.}
  \label{vpaimsrrhsbreakdown}
  \end{minipage}
\hfill
  \begin{minipage}[t]{0.46\linewidth}
{\rotatebox{270}{\mbox{
\includegraphics[width=0.9\textwidth]{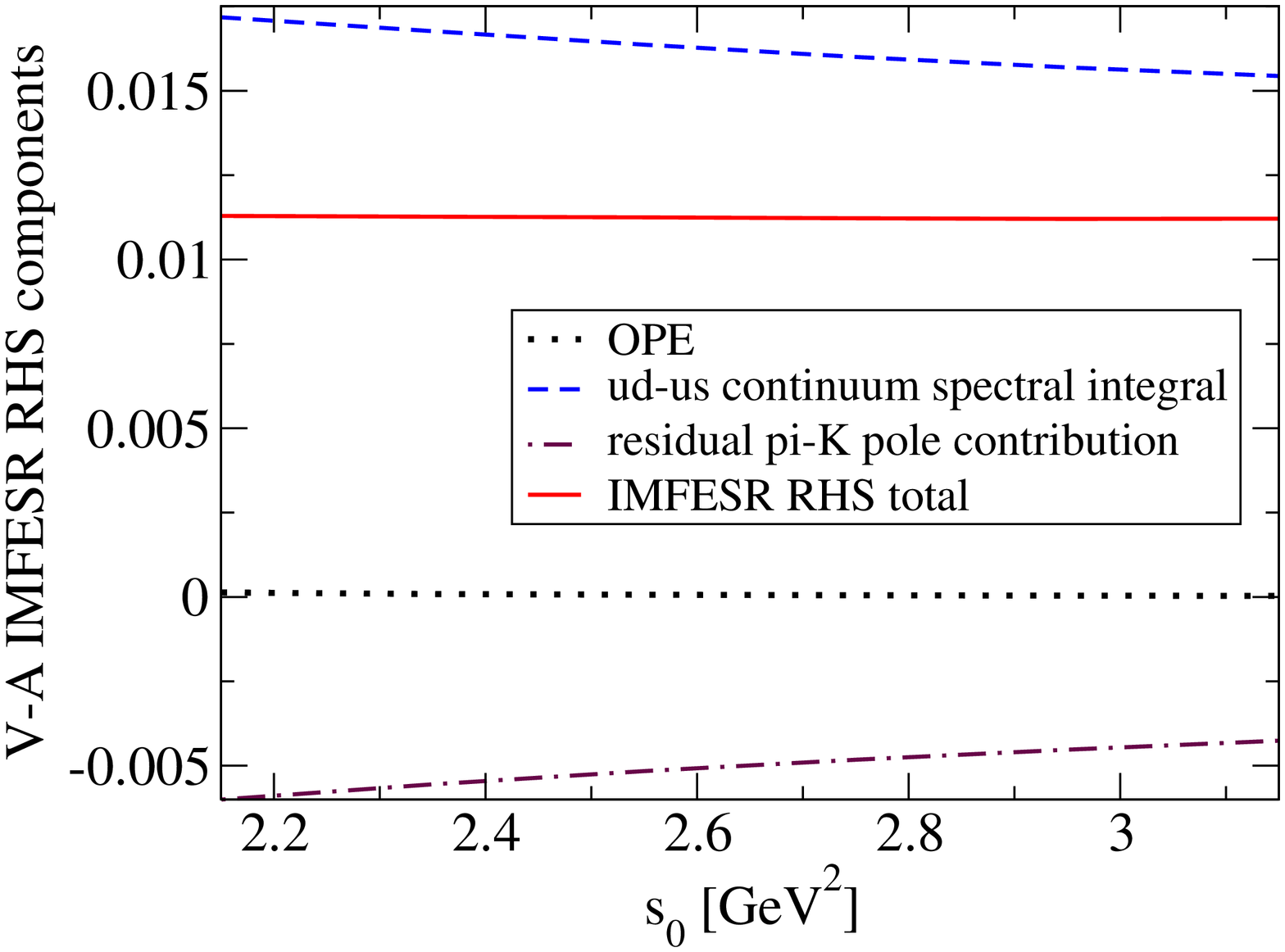}
}}}
   \caption{RHS contributions to the $w_{DK}(y)$ 
$\Delta\overline{\Pi}_{V-A}$ IMFESR, Eq.~(\ref{vpmaimsralt}), 
as a function of $s_0$.}
   \label{vmaimsrrhsbreakdown}
  \end{minipage}
\end{figure}

The dependences on $s_0$ of the OPE, continuum spectral integral and residual
$\pi$/$K$-pole term contributions to the RHSs of the $w_{DK}(y)$
$\Delta\overline{\Pi}_{V+A}$ and $\Delta\overline{\Pi}_{V-A}$ IMFESRs,
Eq.~(\ref{vpmaimsralt}), are shown, for illustration, in 
Figs.~\ref{vpaimsrrhsbreakdown} and \ref{vmaimsrrhsbreakdown}. 
As noted already, OPE contributions are 
negligible for the latter, but not the former. Also shown are the totals 
of all three contributions, which should be independent of $s_0$ and equal 
to $\Delta\overline{\Pi}_{V+A}(0)$ and $\Delta\overline{\Pi}_{V-A}(0)$,
respectively. The $s_0$ stability of these results is obviously
excellent. Similarly good $s_0$ stability is found for the 
$w_{DK}(y)$ $\Delta\Pi_V$ IMFESR and all three $\hat{w}(y)$ IMFESRs.
The corresponding figures are thus omitted for the sake of brevity. Given 
that OPE contributions to the $w_{DK}(y)$ $V-A$ IMFESR are numerically 
negligible, the stability of $\Delta\overline{\Pi}_{V-A}(0)$ with respect 
to $s_0$ for this case supports the treatment of the exclusive $us$ spectral 
integral contributions and $V/A$ separation. The stability in the $V+A$ 
cases tests, in addition, the treatment of the OPE contributions. 
The $s_0$ stability in all cases also supports the neglect of higher
$D$ OPE and residual duality violating contributions in the analysis.

The values for $\Delta\Pi_V(0)$, $\Delta\overline{\Pi}_{V+A}(0)$ and 
$\Delta\overline{\Pi}_{V-A}(0)$ obtained from the $w_{DK}(y)$ analysis are as
follows:
\begin{eqnarray}
&&\Delta\Pi_V(0)\, =\, 0.0230\, (11)_{cont}\, (4)_{OPE}\, (3)_{s_0}
\, =\, 0.0230\, (12)\, ,
\nonumber\\
&&\Delta\overline{\Pi}_{V+A}(0)\, =\, 0.0348\, (10)_{cont}\, (2)_{res}\, 
(5)_{OPE}\, (6)_{s_0}\, =\, 0.0348\, (13)\, ,\nonumber\\
&&\Delta\overline{\Pi}_{V-A}(0)\, =\, 0.0113\, (15)_{cont}\, (2)_{res}\, 
(3)_{OPE}\, (1)_{s_0}\, =\, 0.0113\, (15)\, ,
\label{Q2eq0correlatorresultswdk}\end{eqnarray}
where, to be specific, the central values quoted represent the
average over the $s_0$ window employed. The subscripts 
$OPE$, $cont$ and $res$ identify error components associated with OPE, 
continuum spectral integral and (where present) residual $\pi$/K-pole 
contributions, while the additional component labelled by the subscript 
$s_0$ specifies the small residual variation of the total over the $s_0$ 
analysis window. 

The $\hat{w}(y)$ versions of these analyses, similarly, yield
\begin{eqnarray}
&&\Delta\Pi_V(0)\, =\, 0.0227\, (9)_{cont}\, (6)_{OPE}\, (2)_{s_0}
\, =\, 0.0227\, (10)\, ,
\nonumber\\
&&\Delta\overline{\Pi}_{V+A}(0)\, =\, 0.0348\, (8)_{cont}\, (2)_{res}\, 
(8)_{OPE}\, (3)_{s_0}\, =\, 0.0348\, (12)\, ,\nonumber\\
&&\Delta\overline{\Pi}_{V-A}(0)\, =\, 0.0105\, (11)_{cont}\, (2)_{res}\, 
(5)_{OPE}\, (3)_{s_0}\, =\, 0.0105\, (13)\, .
\label{Q2eq0correlatorresultswhat}\end{eqnarray}

The agreement between the results of
Eqs.~(\ref{Q2eq0correlatorresultswdk}) and (\ref{Q2eq0correlatorresultswhat})
represents a further non-trivial test of the treatment of theoretical and 
spectral integral contributions. The total errors on $\Delta\Pi_V(0)$, 
$\Delta\overline{\Pi}_{V+A}(0)$ and $\Delta\overline{\Pi}_{V-A}(0)$ 
are rather similar for the $w_{DK}(y)$ and $\hat{w}(y)$ determinations,
with spectral integral errors somewhat smaller and OPE errors somewhat 
larger for the $\hat{w}(y)$ case. In view of the similarity of the 
errors, and the fact that the $\hat{w}(y)$ analyses involve both
significantly larger OPE contributions and integrated $D=2$ and $D=4$ OPE 
series having much slower convergence behavior, we take our final results 
to be those obtained from the $w_{DK}(y)$ IMFESRs, whose errors are
dominantly experimental.

The results of Eqs.~(\ref{Q2eq0correlatorresultswdk}), combined with the LEC 
contributions of Eqs.~(\ref{nlonnlolectermcontributions}), the
LEC-independent contributions of Eqs.~(\ref{nonlecudmusrzerovalues}),
and the input values for $L_{5,9}^r(\mu_0)$, yield the final versions
of the IMFESR constraints on the NNLO LEC combinations and (for
$T=V,\, V-A$) $L_{10}^r$. 

We now discuss in more detail the version of this analysis based on the 
weight $w_{DK}$. The $\hat{w}$-based analysis is analogous, and, due to
the good agreement between the results of 
Eqs.~(\ref{Q2eq0correlatorresultswdk}) and (\ref{Q2eq0correlatorresultswhat}),
leads to very similar results for the NNLO LECs. 
These will be displayed, together with those from the $w_{DK}$
analysis, in Table~\ref{table2} below.

The $w_{DK}(y)$ $T=V+A$ IMFESR  becomes
\begin{eqnarray}
C_{12}^r(\mu_0) + C_{61}^r(\mu_0) + C_{80}^r(\mu_0)\, &&=\,
0.00248\, (13)_{cont} (2)_{res} (7)_{OPE} (9)_{s_0} 
(1)_{L_5^r} (8)_{L_9^r} \ {\rm GeV}^{-2}\nonumber\\
&&=\, 0.00248\, (19)\ {\rm GeV}^{-2}\ ,
\label{vpaimfesrlecresult}\end{eqnarray}
where the subscripts $L_5^r$ and $L_9^r$ label errors associated
with the uncertainties on the input $L_{5,9}(\mu_0)$ values and the
labelling of all other sources of error is as specified above.

The $w_{DK}(y)$ $T=V-A$ IMFESR, similarly, yields
\begin{eqnarray}
&&2.12\, L_{10}^r(\mu_0)\, -\, 32(m_K^2-m_\pi^2)\left[ C_{12}^r(\mu_0)
-C_{61}^r(\mu_0)+C_{80}^r(\mu_0)\right]\nonumber\\
&&\qquad\qquad  =\, -0.00346\, (145)_{cont} (15)_{res}
(31)_{OPE} (8)_{s_0} (9)_{L_5^r} (61)_{L_9^r} \nonumber\\
&&\qquad\qquad =\, -0.00346\, (161)\ ,
\label{vmaimfesrlecconstraint}\end{eqnarray}
and the $w_{DK}(y)$ $T=V$ constraint
\begin{equation}
32(m_K^2-m_\pi^2)\, C_{61}^r(\mu_0)\, =\, 0.00727\, (108)_{cont} 
(38)_{OPE} (32)_{s_0} (9)_{L_5^r} (61)_{L_9^r}\, 
-\, 1.06\, L_{10}^r(\mu_0)\ .
\label{vecimfesrlecconstraint}\end{equation}

In Ref.~\cite{boylelatt13vma}, a combined fit incorporating the constraint, 
Eq.~(\ref{vmaimfesrlecconstraint}), the $\Pi_{ud;V-A}(0)$ constraint, 
Eq.~(\ref{udvmaq20constraint}), and lattice $\Pi_{ud;V-A}(Q^2)$ results for 
four $n_f=2+1$, domain-wall fermion RBC/UKQCD ensembles (two with inverse 
lattice spacing $1/a\, =\, 2.31\ {\rm GeV}$ and pion masses $m_\pi\, =\, 293$, 
and $349\ {\rm MeV}$, and two with $1/a\, =\, 1.37\ {\rm GeV}$ and 
$m_\pi\, =\, 171$ and $248\ {\rm MeV}$), was shown to yield 
\begin{eqnarray}
&&L_{10}^r(\mu_0)\, =\, -0.00346(32)\, ,\label{l10rcombined}\\
&&{\cal C}_0^r(\mu_0)\ \, =\, -0.00034(13)\, ,\label{scriptC0combined}\\
&&{\cal C}_1^r(\mu_0)\ \, =\, \ 0.0081(35)\ .\label{scriptC1combined}
\end{eqnarray}
The result (\ref{scriptC0combined}) corresponds to
\begin{equation}
C_{12}^r(\mu_0)-C_{61}^r(\mu_0)+C_{80}^r(\mu_0)\, =\, -0.00055\, (21)
\ {\rm GeV}^{-2}\ .
\label{c12mc61pc80mrho}\end{equation}
Equation~(\ref{l10rcombined}), combined with the $T=V$ constraint,
Eq.~(\ref{vecimfesrlecconstraint}), then implies 
\begin{eqnarray}
C_{61}^r(\mu_0)\, &&=\, 0.00151\, (15)_{cont} (5)_{OPE} (4)_{s_0}
(1)_{L_5^r}\, (8)_{L_9^r}\, (5)_{L_{10}^r} \ {\rm GeV}^{-2}\nonumber\\
&&=\, 0.00151\, (19)\ {\rm GeV}^{-2}\ .
\label{c61value}\end{eqnarray}
There is some correlation between the continuum spectral integral errors and
$L_{10}^r$, but the impact of this correlation does not show up in the 
combined error, to the number of significant figures shown, since
the error on $L_{10}^r$ is strongly dominated by the errors on 
the lattice data.{\footnote{The impact of the uncertainties on $L_5^r$ and 
$L_9^r$ in the $T=V-A$ IMFESR constraint are also very small. 
As an example, doubling the $L_9^r$ uncertainty of 
Ref.~\cite{bt02}, and rerunning the fit of Ref.~\cite{boylelatt13vma},
we find the errors in Eqs.~(\ref{l10rcombined}), 
(\ref{scriptC0combined}) and (\ref{scriptC1combined}) shifted
to $0.00033$, $0.00015$ and $0.0036$ respectively, with no change
in the central fitted values.}}

Taking into account the correlations between the $V$ and $V+A$ 
analysis inputs (or, equivalently, performing the FB $ud-us$ $A$ 
IMFESR analysis directly), one finds, from (\ref{vpaimfesrlecresult}) 
and (\ref{c61value}), 
\begin{eqnarray}
C_{12}^r(\mu_0)\, +\, C_{80}^r(\mu_0)\, &&=\, 0.00097\, (8)_{cont}(2)_{res}
(4)_{OPE} (5)_{s_0} (5)_{L_{10}^r}
\nonumber\\
&&=\, 0.00097\, (11)\ {\rm GeV}^{-2}\ .
\label{c12pc80value}\end{eqnarray}

The determination of $C_{12}^r$ in Ref.~\cite{jop04} has been recently
updated to reflect new values for the main inputs $f_+(0)$
and $f_K/f_\pi$~\cite{mj13}, 
with the result
\begin{equation}
C_{12}^r(\mu_0)\, =\, 0.00005\, (4)\ {\rm GeV}^{-2}\ .
\label{mj13c12}\end{equation}
Eqs.~(\ref{c12pc80value}) and (\ref{mj13c12}) yield
\begin{equation}
C_{80}^r(\mu_0)\, =\, 0.00092\, (12)\ {\rm GeV}^{-2}\ .
\label{mjc12inputc80value}\end{equation}

Replacing the inputs from Eqs.~(\ref{Q2eq0correlatorresultswdk}) with 
those from Eqs.~(\ref{Q2eq0correlatorresultswhat}) and repeating the 
steps just described yields the alternate $\hat{w}$ IMFESR determinations
of the same NNLO LEC combinations shown in Table~\ref{table2}. These
are in excellent agreement with those obtained from the $w_{DK}$ IMFESR 
analysis.

A number of estimates exist in the literature for the three NNLO LECs, 
$C_{12,61,80}^r(\mu_0)$, entering the combinations determined above. 
$C_{61}^r$ was also obtained directly in an earlier version~\cite{dk00} of 
the FB $V$ channel IMFESR analysis,{\footnote{A value for $C_{61}^r(\mu_0)$
differing from that in Table~\ref{table1} was also given in Ref.~\cite{km06}. 
This was meant to represent a translation of the D\"urr-Kambor 
result~\cite{dk00}, which was not given directly in terms of $C_{61}^r$, 
into the explicit $C_{61}^r$ form. The two values turn out to differ 
because of a minor sign transcription error in the translation process. 
Thanks to Bachir Moussallam for clarifying the situation, and tracking 
down the source of the discrepancy.}}
and $C_{12}^r(\mu_0)$ (not determined here) in the lattice analysis of 
Ref.~\cite{bp10} and an updated version~\cite{mj13} of the coupled-channel 
dispersive analysis of Ref.~\cite{jop04}. These estimates/results
are compiled in Table~\ref{table1}. For the quark model results 
of Ref.~\cite{jzlw09}, we quote, for simplicity, the larger of the two 
asymmetric errors from the original publication. In Ref.~\cite{bp10}, a 
number of different results were presented for $C_{12}^r$, corresponding 
to different fit strategies and inputs. Here only the result of Fit IV, 
which did not employ data from the heavier $m_\pi =556\ {\rm MeV}$ ensemble 
and which used updated NLO LEC input (the preliminary version of the results 
of Ref.~\cite{bj11}), has been tabulated. Comparing the quark model and 
RChPT estimates to the IMFESR results above, one sees that the quark model 
does well for $C_{80}^r$ but badly for $C_{61}^r$, while RChPT somewhat 
overestimates $C_{61}^r$ and significantly overestimates $C_{80}^r$.

\begin{table}[H]
\begin{center}
\begin{tabular}{|c|lr|lr|l|}
\hline
LEC&\ \ \ \ RChPT&&\ \  Quark model&&\qquad Other\\
\hline
$C_{12}^r$&-0.00082&\cite{bj11}&-0.00034(2)&\cite{jzlw09}&0.00005(4) 
(Dispersive~\cite{mj13})\\
&-0.00044(16)&\cite{ceekpp05}&&&0.00057(10) \ \ (Lattice~\cite{bp10})\\
&-0.0008(4)&\cite{up08}&&&{}\\
\hline
$C_{61}^r$&\ 0.0021&\cite{km06}&\ 0.00288(26)&\cite{jzlw09}&0.00081(38)
\ \ (IMFESR~\cite{dk00})\\
&\ 0.0019&\cite{abt00}&&&{}\\
\hline
$C_{80}^r$&\ 0.0021(5)&\cite{up08}&\ 0.00087(4)&\cite{jzlw09}&\\
&\ 0.0019&\cite{abt00}&&&{}\\
\hline
\end{tabular}
\end{center}
\caption{Previous results and estimates from the literature for 
$C_{12}^r(\mu_0)$, 
$C_{61}^r(\mu_0)$ and $C_{80}^r(\mu_0)$. LEC values are in units of
${\rm GeV}^{-2}$.}
\label{table1}
\end{table}

An alternate comparison, involving the combinations of NNLO LECs determined 
in the IMFESR analyses above, is given in Table~\ref{table2}. 
Since errors are not quoted for some of the RChPT results in the
literature, we present only central values in this case, using
averages of the different RChPT results listed in Table~\ref{table1}
for each of the $C_k^r$. It is worth noting that the result 
(\ref{c12mc61pc80mrho}) for $C_{12}^r(\mu_0)-C_{61}^r(\mu_0)+C_{80}^r(\mu_0)$ 
differs significantly from the value, $0.00086(67)\ {\rm GeV}^{-2}$, employed in
Ref.~\cite{gapp08}. The difference is due to a combination of two factors:
a significant overestimate of $C_{80}^r$ in the RChPT value used
in Ref.~\cite{gapp08}, and the shift in the $V$ channel IMFESR 
result for $C_{61}^r$ resulting from significant shifts in 
OPE and data inputs.

\begin{table}[H]
\begin{center}
\begin{tabular}{|c||l|l|l|l|}
\hline
LEC combination&RChPT&Quark Model&This work ($w_{DK}$)&This work ($\hat{w}$) \\
\hline
$C_{12}^r+C_{61}^r+C_{80}^r$&\ 0.0034&\ 0.00341(27)&\ \ \  0.00248(19)
&\ \ 0.00248(18)\\
$C_{12}^r-C_{61}^r+C_{80}^r$&-0.0006&-0.00235(25)&\ \ -0.00055(21)
&\ -0.00046(19)\\
$C_{61}^r$&\ 0.0020&\ 0.00288(26)&\ \ \ 0.00151(19)&\ \ 0.00147(17)\\
$C_{12}^r+C_{80}^r$&\ 0.0014&\ 0.00053(2)&\ \ \ 0.00097(11)&\ \ 0.00101(10)\\
\hline
\end{tabular}
\end{center}
\caption{Comparison of quark model and central RChPT estimates 
to the values of the 
NNLO LEC combinations obtained from the various IMFESR analyses above.
LEC combination are understood to be evaluated at $\mu =\mu_0$,
and are in units of ${\rm GeV}^{-2}$.}
\label{table2}
\end{table}

\section{\label{summary}Summary and Discussion}
We have obtained rather good precision determinations of the
NNLO LEC $C_{61}^r$ and NNLO LEC combination $C_{12}^r+C_{61}^r+C_{80}^r$ 
through the use of FB IMFESRs. The much improved low-multiplicity 
B-factory strange hadronic decay distribution data plays an important 
role in achieving the reduced errors, as does the improved
determination of $L_{10}^r$ made possible by the lattice
data on the flavor $ud$ $V-A$ correlator. Our final
results for the NNLO LECs are those given in the previous section.

The determinations based on the $w_{DK}$ and $\hat{w}$ are in excellent 
agreement, and both show good $s_0$ stability. Those based on $w_{DK}$ 
have the additional advantage that the final errors are more dominated 
by their experimental components, and hence less dependent on the 
reliability of the estimates of OPE uncertainties, than are those
based on $\hat{w}$. Because of the strong suppression of high-$s$ 
spectral contributions for the weights employed, the $us$ spectral 
integrals are dominated by contributions from the $K\pi$ mode, which 
has the most accurately measured of the strange exclusive distributions. 
For the $T=V+A$ case, where the $us$ $V/A$
separation uncertainties play no role, the result is that the
errors on the $ud$ continuum spectral integrals (which are a factor
of $\sim 2$ larger than the $us$ continuum integrals) are slightly larger than
the continuum $us$ errors. Improvements to the errors on both the $ud$ and 
$us$ spectral distributions would thus be useful for further reducing the 
errors on our final results. For the $T=V$ and $V-A$ cases, in spite of 
the suppression of contributions from the higher multiplicity
modes, the $us$ $V/A$ separation uncertainty represents the largest component
of the error on the $us$ continuum spectral integrals.{\footnote{As an example,
at the midpoint, $s_0=2.65$ GeV$^2$, of the $s_0$ analysis window, 
the ratio of the $V/A$ separation uncertainty and $K\pi$ distribution error
contributions to the error on the $w_{DK}$-weighted $us$ spectral integral 
is $\sim 1.5$ for the $V$ channel and $\sim 3$ for the $V-A$ channel.}}
The $ud$ continuum errors, however, remain non-negligible, even for the 
$V-A$ case. For the $w_{DK}$-based IMFESRs, there is room for significant 
experimental improvement before reaching the limitations set by the OPE 
uncertainties. Improved $V/A$ separation of the contributions from the 
$K\bar{K}\pi$ and $\bar{K}\pi\pi$ channels can, in principle, be made 
by angular analyses of the B-factory data for these modes, and such 
improvements would serve to significantly reduce the experimental 
components of the errors on the corresponding $T=V$ and $V-A$ IMFESR results.

With regard to the experimental errors,
one should bear in mind that work on the strange distributions and 
branching fractions is ongoing. Preliminary BaBar results
based on the PhD thesis of Adametz~\cite{adametz},
for example, show increases in the branching fractions of the 
$\tau^- \rightarrow K^- \, n\pi^0\nu_\tau $ modes. The dominant
impact of such changes on the current analyses would be through
the normalization of the $K^-\pi^0$ mode contributions, where the preliminary
result $B[\tau^- \rightarrow K^- \pi^0\nu_\tau ]=0.00500(14)$~\cite{adametz}
differs significantly from the current PDG average $0.00429(15)$.
($\bar{K}^0\pi^-$ contributions, whose branching fraction 
normalization is a factor of about two larger, are, however, unaffected.) 
Rerunning the IMFESR analyses discussed above with the preliminary 
$B[\tau^- \rightarrow K^- \, n\pi^0\nu_\tau ]$ results of Ref.~\cite{adametz} 
in place of those used previously and a concomittant adjustment to
the global approximate $ud$ $V,\ A$ rescaling, 
one finds that $C^r_{12}+C^r_{61}+C^r_{80}$ and $C^r_{61}$ are
both shifted downwards by $\sim 1\sigma$, while $C^r_{80}$ is
left essentially unchanged. Explicitly, the results of the $w_{DK}$ 
versions of these modified analyses are
$C_{12}^r(\mu_0) + C_{61}^r(\mu_0) + C_{80}^r(\mu_0)\, =\,
0.00230(18)\ {\rm GeV}^{-2}$,
$C_{61}^r(\mu_0)\, =\, 0.00133(18)\ {\rm GeV}^{-2}$, and
$C_{80}^r(\mu_0)\, =\, 0.00097(11)\ {\rm GeV}^{-2}$.
We stress that BaBar has not yet released their final version of
the analysis of the Adametz thesis data, so, at present, these results 
serve only to illustrate the potential impact of ongoing experimental work. 

We also note that the RChPT estimates for the NNLO LECs considered
here are not quantitatively reliable. This confirms the relevance
of worries expressed elsewhere in the literature about some of the 
aspects of the RChPT approach~\cite{emn10,gp06,mp07}.

Finally, we comment that the result of Ref.~\cite{boylelatt13vma}
for ${\cal C}_1^r(\mu_0)$, which corresponds to 
$C_{13}^r(\mu_0)-C_{62}^r(\mu_0)+C_{81}^r(\mu_0)\, =\, 0.00049(21)
\ {\rm GeV}^{-2}$, provides another example of a $1/N_c$-suppressed LEC 
combination having a non-zero value for $N_c=3$. Such combinations 
are usually neglected in making RChPT estimates, but the non-zero value in 
this case plays a non-trivial role in achieving the improved 
determination of $L_{10}^r$ reported in Ref.~\cite{boylelatt13vma}. 
We also note that the central value for this combination exceeds by a 
factor of $\sim 2.7$ the bound
\begin{equation}
\vert C_{13}^r(\mu_0)- C_{62}^r(\mu_0) + C_{81}^r(\mu_0)\vert
< \vert C_{12}^r(\mu_0)- C_{61}^r(\mu_0) + C_{80}^r(\mu_0)\vert /3
\end{equation}
assumed for it in Ref.~\cite{gapp08}, where the $1/3$ on the RHS 
was meant to reflect the $1/N_c$ suppression of the LHS. This observation 
provides a cautionary note regarding the use of such large-$N_c$ 
assumptions/bounds in contexts where they dominate the errors in 
the full analysis (in the case of Ref.~\cite{gapp08}, that on $L_{10}^r$).

\begin{acknowledgments}
MG is supported in part by the US Department of Energy.
SP is supported by CICYTFEDER-FPA2011-25948, SGR2009-894, and
the Spanish Consolider-Ingenio 2010 Program CPAN (CSD2007-00042). 
KM is supported by a grant from the Natural Sciences and
Engineering Research Council of Canada. KM would also like
to thank Chen Shaomin, Denis Epifanov and Ian Nugent for providing 
details of the ALEPH, Belle and BaBar exclusive $us$ spectral 
distributions, respectively.
\end{acknowledgments}
\appendix
\section{\label{appendix1}OPE Contributions and Errors}
In this appendix we provide details, broken down by dimension
and source, of the total errors on the OPE contributions to
the RHSs of the $w_{DK}$ and $\hat{w}$ $T=V,\, V+A$ and $V-A$ IMFESRs
quoted above. We remind the reader that the OPE terms in question
represent contributions to the IMFESR determinations of the $Q^2=0$ values
of the relevant FB correlator differences, and thus that the relevant
scale for assessing the largeness or smallness of a given contribution is
the corresponding $Q^2=0$ correlator value. To two significant
figures these are, from either Eqs.~(\ref{Q2eq0correlatorresultswdk}) or
(\ref{Q2eq0correlatorresultswhat}), $\Delta\Pi_V(0)=0.023$,
$\Delta\overline{\Pi}_{V+A}(0)=0.035$ and 
$\Delta\overline{\Pi}_{V-A}(0)=0.011$. 

Table~\ref{tableapp1} lists our estimates of the central $D=2$ 
contributions and errors, together with the individual
contributions to these errors. The column headings $\delta m^2$,
$O(\bar{a}^2)$, $prescription$ and $\delta\alpha_s$ label 
individual contributions associated with (i) the uncertainty on the
overall squared mass factors arising from uncertainties in
the FLAG quark mass inputs, (ii) a contribution to the truncation
uncertainty equal to the size of the last ($O(\bar{a}^2$)) term
kept in the truncated series, (iii) the difference between
the results for the three-loop-truncated series obtained using the
central FOPT and alternate CIPT prescriptions, and (iv) the uncertainty
induced by that on the $n_f=5$ $\alpha_s(M_Z^2)$ input, respectively.
We display results only for the smallest and largest $s_0$
employed, $2.15$ and $3.15$ GeV$^2$, respectively. All results
decrease monotonically in magnitude with increasing $s_0$.

\begin{table}[t]
\begin{center}
\begin{tabular}{|ccclcccc|}
\hline
Weight&$T$&$s_0$&$D=2$ integral&$\delta m^2$ &\  $O(\bar{a}^2)$ & 
$prescription$&$\delta\alpha_s$\\
\hline
$w_{DK}$&$V$&2.15&\ \ 0.00106(29)&0.00005&\ \ 0.00011&\ \ 0.00027&\ \ 0.00002\\
&&3.15&\ \ 0.00061(13)&0.00003&\ \ 0.00005&\ \ 0.00012&\ \ 0.00001\\
&$V+A$&2.15&\ \ 0.00211(59)&0.00011&\ \ 0.00022&\ \ 0.00053&\ \ 0.00003\\
&&3.15&\ \ 0.00121(26)&0.00006&\ \ 0.00010&\ \ 0.00023&\ \ 0.00001\\
&$V-A$&2.15&\ -0.00001(1)&0.00000&\ \ 0.00000&\ \ 0.00000&\ \ 0.00000\\
&&3.15&\ -0.00000(0)&0.00000&\ \ 0.00000&\ \ 0.00000&\ \ 0.00000\\
\hline
$\hat{w}$&$V$&2.15&\ \ 0.00196(44)&0.00010&\ \ 0.00038&\ \ 0.00020&\ \ 0.00004\\
&&3.15&\ \ 0.00109(20)&0.00006&\ \ 0.00018&\ \ 0.00007&\ \ 0.00001\\
&$V+A$&2.15&\ \ 0.00391(88)&0.00020&\ \ 0.00075&\ \ 0.00040&\ \ 0.00007\\
&&3.15&\ \ 0.00219(40)&0.00011&\ \ 0.00036&\ \ 0.00015&\ \ 0.00003\\
&$V-A$&2.15&\ \ 0.00001(1)&0.00000&\ \ 0.00001&\ \ 0.00000&\ \ 0.00000\\
&&3.15&\ \ 0.00001(0)&0.00000&\ \ 0.00000&\ \ 0.00000&\ \ 0.00000\\
\hline
\end{tabular}
\end{center}
\begin{quotation}
\caption{\label{tableapp1}The $w_{DK}$ and $\hat{w}$ IMFESR $D=2$ 
OPE assessments, total errors and error components for the 
$T=V,\, V+A$ and $V-A$ channels and $s_0=2.15$ and $3.15\ {\rm GeV}^2$.
The $s_0$ entries are in GeV$^2$ and the error components are
labelled as described in the text.}
\end{quotation}
\end{table}%
From the Table we see that $D=2$ contributions are entirely negligible 
for $T=V-A$. The central $D=2$ OPE contributions are also small,
though not negligible, for the other channels, varying, for example
for $w_{DK}$, from $5\%$ to $3\%$ of $\Delta\Pi_V(0)$ for $T=V$ and 
$6\%$ to $4\%$ of $\Delta\overline{\Pi}_{V+A}(0)$ for $T=V+A$, as $s_0$ 
is increased from $2.15$ to $3.15$ GeV$^2$. The corresponding total $D=2$
errors, similarly, vary from $1\%$ to $0.6\%$ of $\Delta\Pi_V(0)$ and
$2\%$ to $0.7\%$ of $\Delta\overline{\Pi}_{V+A}(0)$ over the 
same range. The prescription dependence is the dominant contribution
to the total error for $w_{DK}$, while both the prescription dependence
and $O(\bar{a}^2)$ truncation error contribution play a significant
role for $\hat{w}$. The $D=2$ errors for $\hat{w}$ are $\sim 50\%$
larger than those for $w_{DK}$.

Table~\ref{tableapp2} contains our $D=4$ contributions and total
errors. The errors are the quadrature sum of (i) the uncertainty generated
by that on the input FLAG ratio of strange to light quark masses,
(ii) a truncation uncertainty equal to the last ($O(\bar{a}^2)$) term
kept in the truncated $D=4$ series, and (iii) the uncertainty
generated by that on $r_c$. Since the $r_c$-induced uncertainty
is much larger than the other two, we quote only the total error
in this case.
\begin{table}[t]
\begin{center}
\begin{tabular}{|cccl|}
\hline
Weight&$T$&$s_0$&\ \ $D=4$ integral\\
\hline
$w_{DK}$&$V$&2.15&\ \ \ 0.00028(7)\\
&&3.15&\ \ \ 0.00013(3)\\
&$V+A$&2.15&\ \ \ 0.00051(15)\\
&&3.15&\ \ \ 0.00024(7)\\
&$V-A$&2.15&\ \ \ 0.00006(2)\\
&&3.15&\ \ \ 0.00002(1)\\
\hline
$\hat{w}$&$V$&2.15&\ \ \ 0.00173(39)\\
&&3.15&\ \ \ 0.00080(18)\\
&$V+A$&2.15&\ \ \ 0.00270(76)\\
&&3.15&\ \ \ 0.00129(36)\\
&$V-A$&2.15&\ \ \ 0.00077(32)\\
&&3.15&\ \ \ 0.00030(12)\\
\hline
\end{tabular}
\end{center}
\begin{quotation}
\caption{\label{tableapp2}The $w_{DK}$ and $\hat{w}$ IMFESR $D=4$ 
OPE estimates and total errors for the
$T=V,\, V+A$ and $V-A$ channels and $s_0=2.15$ and $3.15\ {\rm GeV}^2$.
The $s_0$ entries are in GeV$^2$.}
\end{quotation}
\end{table}%
$D=4$ errors for the $V$ and $V+A$ channels are much smaller than the
corresponding $D=2$ errors for $w_{DK}$, but grow to
$\sim 90\%$ of the corresponding $D=2$ errors for $\hat{w}$. The
$D=4$ contributions are also sub-leading ($\sim 20-25\%$ of the
$D=2$ ones) in the $V$ and $V+A$ channels for $w_{DK}$. For $\hat{w}$,
in contrast, they range from $88\%$ to $73\%$ and $69\%$ to $58\%$ of the 
$D=2$ contributions for the $V$ and $V+A$ channels respectively. $D=4$
$V-A$ contributions, though larger than the strongly suppressed
$D=2$ ones, are still very small for $w_{DK}$, and do
not exceed $7\%$ of $\Delta\overline{\Pi}_{V-A}(0)$ for $\hat{w}$.

As noted in the text, a very conservative error, equal to the value
of the $ud$ contribution to the FB difference, is employed 
for the sum of the FB $D=6$ and $8$ contributions. The central value
is obtained using the VSA for the $D=6$ contributions and setting
$D=8$ contributions to zero. The $ud$ contribution used to set the
error on this (very small) central value
is evaluated for $T=V$ and $V+A$ using the fit values
for $C_6^V$, $C_6^A$, $C_8^V$ and $C_8^A$ obtained in Ref.~\cite{dv72},
and for $T=V-A$ using the direct fits for the $V-A$ channel
analogues, $C_6^{V-A}$ and $C_8^{V-A}$, obtained in Ref.~\cite{bgjmp13}.
The resulting central $ud-us$ $D=6+8$ estimates, together with
the $ud$ $D=6+8$ contributions and their errors (the latter generated
by the errors and correlations on the fitted $D=6$ and $8$ coefficients) 
are listed in Table~\ref{tableapp3}.
\begin{table}[t]
\begin{center}
\begin{tabular}{|ccccl|}
\hline
Weight&$T$&$s_0$&Central $ud-us$
&\ \ $ud$ $D=6+8$ \\
&&& $D=6+8$ integral\ \ &\ \ \ \ integral\\
\hline
$w_{DK}$&$V$&2.15&-0.00000&\ \ 0.00048(20)\\
&&3.15&-0.00000&\ \ 0.00013(5)\\
&$V+A$&2.15&\ 0.00000&\ \ 0.00011(47)\\
&&3.15&\ 0.00000&\ \ 0.00002(13)\\
&$V-A$&2.15&-0.00001&\ \ 0.00045(17)\\
&&3.15&-0.00000&\ \ 0.00013(4)\\
\hline
$\hat{w}$&$V$&2.15&\ 0.00001&\ -0.00053(22)\\
&&3.15&\ 0.00000&\ -0.00017(7)\\
&$V+A$&2.15&-0.00001&\ \ 0.00001(51)\\
&&3.15&-0.00000&\ \ 0.00000(16)\\
&$V-A$&2.15&\ 0.00002&\ -0.00066(11)\\
&&3.15&\ 0.00001&\ -0.00021(4)\\
\hline
\end{tabular}
\end{center}
\begin{quotation}
\caption{\label{tableapp3}$s_0=2.15$ and $3.15$ GeV$^2$ values
of the estimated FB $D=6+8$ contributions, together with the
flavor $ud$ $D=6+8$ OPE contributions and errors used to
set the uncertainty on the estimated central values, for the 
$T=V,\, V+A$ and $V-A$ channel versions of the $w_{DK}$ and $\hat{w}$ 
IMFESRs. The $s_0$ entries are in GeV$^2$.}
\end{quotation}
\end{table}%
For $T=V+A$ there are strong cancellations between the separate
$V$ and $A$ contributions, with the result that the central value
of the $D=6+8$ $ud$ $V+A$ sum is much smaller than the corresponding
uncertainty. No such strong cancellation occurs in either 
of the $V$ or $V-A$ channels. To maintain our $D=6+8$ bound as a 
conservative one for all three cases, we have thus taken as the 
final versions of the error bounds on the FB $ud-us$ $D=6+8$ 
contributions, the sum of the absolute values of the corresponding 
central $ud$ contribution and its error. These can be read off
directly from the results quoted in the Table. The resulting
$D=6+8$ error is the largest of the OPE error 
components for the $V$ and $V-A$ channels, and non-negligible,
but somewhat smaller than the $D=2$ error, for $V+A$.


\vfill\eject
\end{document}